\documentclass[letterpaper]{IEEEtran}
\IEEEoverridecommandlockouts

\input{glyphtounicode}
\usepackage{microtype}

\usepackage{flushend}

\usepackage{amsmath,amssymb,amsthm}
\DeclareMathSizes{6.6}{6.6}{4.62}{5} % the class's 6.6pt title-block font: no 3.3pt math fonts

\usepackage{graphicx}
\graphicspath{{./figures/}}
\usepackage{array,booktabs}
\usepackage{stfloats}
\usepackage{xcolor}
\usepackage{textcomp}

\usepackage{algorithm}
\usepackage{algpseudocode}
\floatstyle{ruled}\restylefloat{algorithm}
\algrenewcommand\algorithmicrequire{\textbf{input:}}
\algrenewcommand\algorithmicensure{\textbf{output:}}
\algrenewcommand\algorithmiccomment[1]{\hfill$\triangleright$~#1}

\usepackage{cite}
\makeatletter
\mathchardef\citeprepenalty=\@M % a citation never starts a line
\makeatother
\usepackage{xurl}

\usepackage{hyperref}
\hypersetup{
  hidelinks,
  pdftitle={Halley: A Predictive Control Plane for Optical AI Data Center Networks},
  pdfauthor={Manav Mishra and Krishna M. Sivalingam},
  pdfsubject={Arxiv},
  pdfkeywords={co-packaged optics, common-cause failure, data center networks,
    distributed training, microring modulators, optical circuit switching,
    optical interconnects, predictive control, reliability},
  pdfdisplaydoctitle=true,
  bookmarksnumbered=true,
  bookmarksopen=true,
  bookmarksopenlevel=1
}

\newtheorem{theorem}{Theorem}
\newtheorem{corollary}{Corollary}
\newtheorem{proposition}{Proposition}
\newtheorem{observation}{Observation}

\makeatletter
\renewcommand{\p@subsection}{\thesection\mbox{-}}
\renewcommand{\p@subsubsection}{\thesection\mbox{-}\thesubsection\mbox{-}}
\makeatother

\newcommand{\Ebb}{\mathbb{E}}
\newcommand{\Var}{\operatorname{Var}}
\newcommand{\Beta}{\mathrm{Beta}}
\newcommand{\Fbeta}{F_{\mathrm{Beta}}}
\newcommand{\Astar}{A^{*}}
\newcommand{\rhobar}{\bar{\rho}}
\newcommand{\degC}{\textdegree C}

\begin{document}

\title{Halley: A Predictive Control Plane for Optical
  AI Data Center Networks}

\author{Manav Mishra$^1$
    and Krishna M. Sivalingam$^2$ \\
$^1$\textit{Prompeteer.ai, Los Angeles, CA
  90066, USA}  \\
$^2$\textit{Department of Computer Science and Engineering,
  Indian Institute of Technology Madras, Chennai, India}
  \\
  Emails: \texttt{manav@prompeteer.com, skrishnam@cse.iitm.ac.in} \\
  CORRESPONDING AUTHOR: Manav Mishra.
}

\maketitle

\begin{abstract}
A frontier AI model is trained in an AI data center, on tens of
thousands of accelerators joined by a dedicated network, and every
training step ends with all of them exchanging data at once. Between
racks, that network is optical, and each of its four layers, from lasers
to switches, runs a fixed or after-the-fact policy. Lasers sharing an
enclosure, cooling, and power can fail together. At a fixed spare ratio,
correlated failure provably caps a pool's availability however large it
grows.  In this paper, we present a comprehensive predictive control
plane framework, called Halley, that reads each resource's next state
and acts before the failure occurs. It has four components, each
carrying a formal guarantee: LightReserve that sizes the laser pool to
provide a coverage bound, FoldFit that schedules the switch within
$11/9$ of the optimal, ThermaLead that holds a microring with
one-step-ahead control, and TriageGuard that protects the highest-risk
modules first.  Halley gives each layer
a predictive controller that defers to the deployed policy when
unsure. Composed from simulations at that point, effective goodput, the
share of ideal throughput kept, rises from 65.2\% to 76.5\% if a window
that exhausts its spares is lost. If only the dark channels' bandwidth
is lost, it reaches 67.4\%. Per megawatt, that is up to 1.03--1.17 times
the training; for the same training, up to 3.3--14.7\% less data center
capacity. \textsf{Halley}'s fallback circuits are formally verified and are shown
to achieve close timing on a commercial FPGA, and pass a 130~nm signoff.
\end{abstract}

%% At a stress-level failure probability of 0.05 per
%% restock window and correlation 0.15, the cap for one spare per active
%% laser is 99.670\%. At the industry's 50~FIT target, 99.999\% still takes
%% over ten times the spares of independent failures. Separating lasers
%% into thermal and power domains first reaches 99.999\% at under a fifth
%% the cost of spares alone, under our cost model.

 \begin{IEEEkeywords}
Co-packaged optics, common-cause failure, data center networks,
distributed training, microring modulators, optical circuit switching,
optical interconnects, predictive control, reliability
 \end{IEEEkeywords}

\section{Introduction}
\label{sect:intro}

A frontier AI training run usually occupies a single cluster in an AI
data center, where a dedicated network moves data among tens of
thousands of accelerators.  Between racks, copper cannot carry that data
within its power and reach budget, so the network is optical
\cite{ref1}, and the industry is now moving the optics onto the package
itself, next to the switch or accelerator silicon, an arrangement known
as co-packaged optics (CPO).

This network, the fabric, has an optical data plane of four layers, each
managed on its own today. A shared pool of external lasers supplies
light to many photonic engines, the on-package optical transceivers of
CPO. An optical circuit switch (OCS) joins those engines to
one another by connecting whole light paths instead of forwarding
individual packets. Microring modulators, ring resonators a few microns in radius, encode the data onto each wavelength. The lasers arrive in
field-replaceable modules that are swapped in service. 

At hundred-thousand-accelerator scale, reliability, network topology,
and checkpointing have become first-order design constraints
\cite{ref1}. Each
layer faces an online allocation decision under a hard service
constraint: how many spare lasers the pool should hold, which circuits
the switch should hold and for how long, where each ring's heater should
be set, and which modules receive the few hot spares, each wired to its
module in advance. The deployed policies this paper replaces, which we
call reactive for short, either fix each answer once in advance or make
it only after the event it should have anticipated. Training is
bulk-synchronous: every training step ends in a collective, a phase in
which all accelerators communicate at once and which also acts as a
barrier, so the step finishes only when its slowest link does. A drifted
ring, a failed laser, or a late switch reconfiguration therefore leaves
every accelerator idle at that barrier.

Software-defined networking provisioned for a separate control plane
from the data plane.  Production optical fabrics use that approach for
provisioning. For example, Google's Jupiter fabric routes on a predicted
traffic matrix \cite{ref13} and TPUv4 reconfigures its optical circuit
switches to ride through outages \cite{ref34, nr_jouppi23}.  Similarly,
Ironwood TPU's 9,216-chip superpods use them to route around
interruptions, the MixNet research system reconfigures ahead of a
predicted all-to-all \cite{ref14}. The Opus system reconfigures photonic
rails at the boundaries between parallelism phases
\cite{nr_ding26}. Failure prediction for pluggable optical transceivers
is now validated on production data \cite{nr_su26, nr_wang26b}. \textit{We
observed that existing control planes do not forecast the co-packaged
optical hardware itself, a laser's failure or a ring's drift, and hence
do not act before the failure or drift arrives.}

\noindent\textbf{Proposed Halley framework:} This work presents
\textsf{Halley}, a control plane that provides these necessary
features. The system has multiple components as described below. The
\textsf{Halley} framework provides every layer with a component of its
own. \textit{LightReserve} sizes the laser pool to a coverage bound
instead of a fixed ratio: a floor on the chance that its spares absorb
every failure in a window, valid for the assumed failure rate and
correlation. \textit{FoldFit} schedules the OCS so that, under stated
conditions, a schedule's total dwell time, the time its circuits are
held, stays within $11/9$ of the shortest possible, a bound that leaves
out switching delay.  \textit{ThermaLead} sets each microring's heater
ahead of the drift. \textit{TriageGuard} steers a limited budget of hot
spares to the modules most likely to fail. Each component predicts its
layer's next state and acts before it arrives. Each steps aside to the
reactive system that it replaces when its forecast is stale or
unsure. Hence, for such a forecast the result is no worse than current
systems, though a confident but wrong forecast has no such floor yet.

Adding spares is the usual way to keep a shared laser pool lit. In a
co-packaged pool the lasers are unlikely to fail independently. NVIDIA's
external laser module holds eight lasers in one temperature-controlled
enclosure \cite{nvels}. Hence, the lasers of a module share its heat and
often one power supply, and a thermal or power fault that stresses one
laser stresses its neighbors as well. Failures that arrive together can
exhaust the spares. Growing the pool at the same spare ratio, the share
of the pool held as spares, does not spread that risk away, since the
next burst of failures is correlated too.

At a fixed spare ratio, availability converges as the pool grows to a
closed-form ceiling $\Astar < 1$ that depends on the failure correlation
$\rho$ and the per-laser failure probability but not on the pool's
size. Raising the spare ratio does lift availability, but past a certain
point so slowly that a high target needs a pool larger than any operator
would deploy. The limit itself is classical. When exchangeable failures
are positively correlated, the failed fraction of a large pool converges
to a random level instead of to its mean \cite{aldous85}. Common-cause
models in reliability engineering bound redundant systems the same way
\cite{nureg5485, iec61508-6}. What this paper adds is its consequence
for a pooled laser supply: where the cap binds, what it costs in spares,
and the cheaper alternative. However, correlation can be
controlled. Separating the lasers into distinct thermal and power
domains and sourcing them from diverse batches raises the ceiling, which
redundancy cannot.

\noindent\textbf{Technical contributions:} The main contributions are:

\begin{itemize}

  \item First, it carries the classical exchangeable limit over to
    pooled laser supplies and proves that in a pool of many modules the
    correlation between modules, more than within them, sets the
    cap. Under a stated cost model, it also finds that decorrelating
    before buying spares is the cheaper path across the range of
    correlations considered plausible.

\item Second, it builds the proposed \textsf{Halley} control plane
  framework, that has four major components.  Three of its four
  components carry a formal guarantee, stated in idealized form later.
  The novel aspect is the design for the failure physics of the AI data
  center.

  \item Third, it composes the components into effective cluster goodput
    under two readings of a spare shortfall, couples the laser and
    thermal losses through a shared common cause, reports the tail as
    well as the mean, and proves a closed form for the
    correlation-driven part of that tail.

  \item Fourth, it synthesizes, places, and routes the fallback circuits
    for a commercial field-programmable gate array, carries the same
    sources through signoff as a SkyWater 130~nm layout, and formally
    verifies their register-transfer design, the part of \textsf{Halley} that can
    be built today.  

\end{itemize}

The paper also reports a detailed performance study of the proposed
mechanisms, based on simulation and mathematical modeling, and their
circuit-level realization: an FPGA build and a signed-off SkyWater 130~nm
layout. The software implementation of these components and the hardware
design are available at \cite{implzip26}.

\noindent\textbf{Important Results:}
At the correlation we assume for a well-built
package and the stress failure rate of Section~\ref{sect:analysis}, a
pool with one spare per active laser, the $2\times$ budget cap we assume
throughout, cannot pass 99.670\% availability however large it grows. A
reserve of four spares for 32 channels, the proportion an analyst
reports for a related shipping part, reaches only 86.3\%. Availability
of 99.999\% is then out of reach for any spare ratio within that cap.
For a 32-channel pool at that correlation, 99.999\% stays out of reach
within the $2\times$ cap whenever a laser's failure probability per
restock window exceeds about $2.4 \times 10^{-4}$. At 50 failures per
billion hours (50~FIT), the Optical Internetworking Forum (OIF) target
for a laser source, which we apply to each laser, a laser crosses that
probability once the restock window exceeds about 200 days.

%% Correlation also raises the price of 99.999\% availability. At the five
%% rates of Table~\ref{tab:regime}, from the 50~FIT target to the stress
%% rate, it takes 16 to 25 times the spares that independent failures would
%% need.

With the spare count held at the 23 lasers our sizing rule returns for a
99\% target, cutting the correlation from $\rho = 0.15$ to $\rho = 0.03$
moves availability from about 99\% to just past 99.999\%. Under a stated
cost model at the stress rate, reaching 99.999\% by separating the pool
first and then sizing its spares costs \$44,100 in all, against
\$234,000 for spares alone.

The overall figure of merit is effective cluster goodput, defined as the
fraction of the cluster's ideal throughput that survives the losses of
the fabric. At the stress setting, composing the simulated components
lifts it from 65.2\% to 76.5\%, a 17.3\% relative gain. That figure
counts every window whose failures exhaust the spares as lost, which we
call the coverage reading. On the lit-share reading, in which only the
dark channels' bandwidth is lost, goodput rises from the same 65.2\%
baseline to just 67.4\%. Most of the gain thus comes from a provisioning
decision: the larger spare pool that LightReserve sizes for the
correlation a static reserve ignores. Holding that pool to the $2\times$
cap costs only 0.2 points, and where no spare count within the cap
reaches a target, LightReserve determines the decorrelation it
needs. ThermaLead's simple forecast does slightly worse than the
deployed dither lock on the stress trace; gated to fall back to that
lock whenever its recent error is the larger of the two, it has 2.1\%
fewer lock losses. At 50~FIT and a 90-day restock the fixed reserve
already covers nearly every window, so the composed relative gain falls
to about 1.3\%; with yearly restocks, 99.999\% still exceeds the
$2\times$ cap at 50~FIT.

Within each reading, the stress-case gain is an upper bound on the
useful compute freed from the same silicon and power, because it treats
fabric stall as the only recoverable loss.  This bound matters because
power and local approval now limit how fast new AI data centers are
built. A new grid connection can take five to ten years
\cite{iea26keyquestions}. Around 45 US data center projects worth
\$68~billion were blocked or delayed at the local level between April
and June 2026 \cite{purnell26dcwatch}. Every point of goodput that the
fabric loses is capacity that was built and powered but did no useful
work. At the bound, what \textsf{Halley} recovers means up to 14.7\%
fewer new megawatts to site and connect for a given training target, or
up to 1.17 times as much training from the power already in place, with
no wait for a new grid connection.  On the lit-share reading the figures
are 3.3\% and 1.03 times. Energy per unit of training falls too, though
by less, since a waiting accelerator draws less than full power.
%\cite{\ref24}

%% If the common cause that caps laser availability also drives the thermal
%% loss at the microrings, as the coupled model of
%% Section~\ref{sect:corfaccomp} assumes, multiplying the two as
%% independent understates the risk: on the coverage reading the mean
%% goodput stays at 76.5\%, but the fifth percentile is 67.8\%.

The remainder of the paper is organized as
follows. Section~\ref{sect:bg} presents the background material and
related work. Section~\ref{sect:ceil} presents the impact of failure
correlation on various metrics such as
availability. Section~\ref{sect:cpfrwk} presents the control plane
framework and the four mechanisms proposed in this
paper. Section~\ref{sect:perf} presents a detailed performance study of
the proposed mechanisms based on simulation and mathematical modeling
studies. Section~\ref{sect:hwimp} presents the circuit-level realization
of the proposed mechanisms, built for a commercial FPGA and signed off as
a SkyWater 130~nm layout. Section~\ref{sect:disc} presents a detailed discussion of
the proposed framework including its limitations.
Section~\ref{sect:conc} concludes the paper.

\section{Background}
\label{sect:bg}

This section presents the relevant background and related material.

%% Each figure and table caption ends with a provenance tag,
%% defined in Section~\ref{sect:simmeth}.

\begin{figure*}[!t]
\centering
\includegraphics[width=\linewidth]{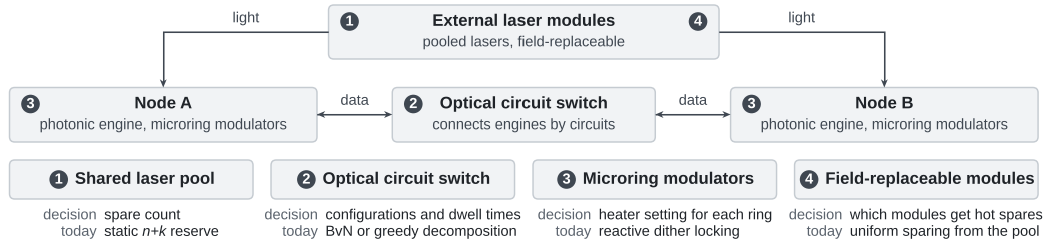}
\caption{The CPO data plane. Top: its four layers, each with the decision it must make and the policy deployed today. Bottom: the light path. External laser modules supply light to each photonic engine, microring modulators in the engine encode data onto it, and the OCS switches the modulated signals between engines [schematic].}
\label{fig:cpo}
\end{figure*}

\subsection{Data Center Networks}

The fabric is kept separate from the general-purpose front-end network,
and purpose-built production fabrics such as Alibaba's High Performance
Network (HPN) have moved far from the classical data center architecture
\cite{ref33}.

\subsubsection{Optical Networking}

Optics replaces a wiring problem with an allocation problem at two
levels: inside a single link and across the fabric that joins the links.
Inside a link, wavelength division multiplexing (WDM) multiplies capacity without adding cable, because
each channel travels on its own wavelength, and that makes capacity
allocation a scheduling problem: which transmitter sends on which
wavelength, and when. The problem has a long literature on star-coupled
and mesh WDM networks, covering prediction-based and interval-based
schedulers \cite{ref5, ref6, ref7} and restoration mechanisms for
channel and link failures \cite{ref8, ref9}.

The same problem recurs across the fabric in optical circuit switching.
An OCS passes the signal through untouched at any line rate, but demand
shifts, so a scheduler must break the demand matrix, the traffic each
endpoint must send to each other endpoint, into a sequence of switch
configurations.

\subsubsection{Training Traffic}

Data, tensor, and pipeline parallelism split a training job across the
accelerators. Mixture-of-experts (MoE) models add expert parallelism,
which spreads specialist parts of the model, the experts, across the
job's worker processes and routes each token to a few of them
\cite{ref14, ref37}. Expert dispatch is an all-to-all collective. Its
demand matrix is dense within each expert group and sparse across
groups, and it shifts as some experts draw more tokens than others.
FoldFit schedules these matrices in Section~\ref{sect:cpfrwk}.

Published runs show the scale: Llama~3 was trained on about 16,000 GPUs
\cite{ref24}, MegaScale scaled training past 10,000 \cite{ref25},
DeepSeek-V3 used 2,048 \cite{ref36}, and Meta reports training on the
order of 100,000 GPUs \cite{nr_salpekar26}. At this scale hardware is
the main threat. Llama~3 attributed about 78\% of its unexpected
interruptions to confirmed or suspected hardware issues \cite{ref24}.
 A study of two large research
clusters reports a mean time to failure of 7.9 hours for 1,024-GPU jobs;
InfiniBand link failures were among the top causes of job failures, and
most of those came in one short burst from a handful of nodes
\cite{ref35}. Meta's report on a 32,000-GPU job counts 2.3 interruptions
per 1,000 servers per day, 78\% of them hardware-related, and Meta
estimates one failure every 18 minutes at 100,000 GPUs
\cite{nr_salpekar26}.

\subsection{Optical Data Plane in a Data Center Network}
\label{sect:odp}

Fig.~\ref{fig:cpo} shows the four optical layers of the data plane.

\paragraph{The CPO data plane} CPO moves the photonic engines from the
rack edge onto the switch or accelerator package, and the light source
changes with them: a shared pool of external lasers, in
field-replaceable modules, replaces one laser per engine. Each NVIDIA
Quantum-X package uses 18 photonic engines fed by 36 laser inputs
\cite{ref11}. The light comes from external laser modules of eight
lasers each, and one module powers 32 of the switch's 576 transmit lanes
\cite{nvels}, which implies 18 modules and 144 lasers per switch; the 36
laser inputs above describe a single package. The OIF implementation
agreement for external laser modules, the external laser small
form-factor pluggable (ELSFP), allows an optional thermoelectric cooler
and replacement at the faceplate \cite{np_oif25}. An analyst reports
that each related Spectrum-X package carries 36 photonic engines, 32
active and four held in reserve for redundancy \cite{ref11}; NVIDIA's
own description of the single-ASIC version lists 32 engines fed by 16
laser modules and mentions no reserve \cite{nvels}. We apply that
proportion of engines to the laser pool, so 32 active channels backed by
a four-laser reserve is this paper's calibration point throughout, and
every later use of the four-laser reserve rests on this
assumption. Other vendors also use external lasers. Broadcom's
Tomahawk~6-Davisson, sampling since October 2025, places 102.4~Tb/s of
CPO behind field-replaceable ELSFP laser modules. Light sources are also
changing form. Lightmatter's Guide is a one-rack-unit light engine whose
output would otherwise take about 18 ELSFP modules
\cite{np_lightmatter26a}. Lumentum has announced an eight-wavelength
dense-WDM ELSFP with initial availability in the first half of 2027
\cite{np_lumentum26b}. How many lasers share an enclosure, the grouping
behind the module-level correlation of Section~\ref{sect:hier},
therefore varies by design. The alternative keeps the optics in a
pluggable module that can be serviced on its own.  Mesh Optical's
linear-pluggable optics drop the digital signal processor that retimes
and equalizes the signal, and the vendor reports a draw under 10~watts
at 1.6~Tb/s \cite{ref43}. NVIDIA, which reported Spectrum-X Photonics in
full production in June 2026, presents it as giving five times the power
efficiency and five times the AI uptime of networks built on pluggable
transceivers \cite{np_nvidia26a}.

\paragraph{Optical circuit switching and MoE traffic} Jupiter brought
optical circuit switching to data center scale under software-defined
control \cite{ref13, nr_liu23}, and Sirius to nanosecond switching
\cite{ref15}.  An earlier system, Helios, paired electrical packet
switching with optical circuit switching in modular data centers and
reconfigured the circuits as traffic shifted \cite{farrington}.  Other
work ties the topology to the training job. TopoOpt optimizes the
topology together with the parallelization strategy \cite{ref31}, and
MixNet exploits the locality of expert parallelism in MoE training,
beating a topology optimizer by $1.5\times$ and $1.3\times$ in its
reported comparisons \cite{ref14}. For scheduling, circuit switches
classically use a Birkhoff--von Neumann (BvN) decomposition
\cite{ref16}, which meets the busiest-port lower bound on the makespan,
the total time the schedule holds its circuits, but does so by splitting
demands across as many as $(P-1)^2+1$ configurations for $P$ ports, each
paying a reconfiguration delay. Faster heuristics, such as MixNet's
greedy reconfiguration \cite{ref14}, state no worst-case bound. Research
schedulers model the cost of switching explicitly. Solstice schedules
hybrid circuit and packet switches by heuristic and lands within 14\% of
optimal at scale, without a worst-case bound \cite{solstice}. Eclipse
carries a constant-factor guarantee on the traffic a circuit switch
serves within a fixed time window \cite{eclipse}, and Sunflow finishes a
coflow, a group of flows that completes together, within $2\times$ of
the optimal time \cite{sunflow}. Solstice and Eclipse halt the whole
switch for a delay $\delta$ at every change of configuration, and
Sunflow halts only the ports being reconfigured. Work from 2025 and 2026
schedules circuit switches for machine-learning jobs
specifically. Chronos derives circuit schedules ahead of time from the
training program, leaving expert parallelism as the one pattern it must
schedule on demand \cite{nr_renganathan25}.  Spectra schedules one AI
demand matrix across parallel circuit switches with the reconfiguration
delay in the objective \cite{nr_liang26a}, LACE and Flux compile
reconfiguration schedules from the structure of the whole workload
\cite{nr_liang26b, nr_troch26}, DELTA engineers one static topology from
the job's dependency graph \cite{nr_ye26a}, OMLSched schedules jobs on
optical ML clusters \cite{nr_hu26a}, and Silverstone reconfigures the
fabric across the phases of disaggregated RL \cite{nr_tan26}. Amponsah
and Addanki show that the many matchings of a BvN decomposition fragment
MoE execution into small batches \cite{nr_amponsah26}. Coflow schedulers
for parallel and multi-core circuit switches now carry worst-case
guarantees with the reconfiguration delay in the model \cite{nr_chen25,
  nr_wang26a}; Chen's makespan ratio grows with the number of flows at a
port. Section~\ref{sect:ff} sets out how the decomposition works, what
it costs, and how FoldFit's bound compares with these.

\paragraph{Microring stabilization and module reliability} A microring
modulator responds to light only near one particular wavelength, its
resonance, and that wavelength moves with temperature. A standard 220-nm silicon ring's resonance shifts about 77~pm/\degC{}
near 1550~nm \cite{xu14}, and our model uses 70~pm/\degC{} as a representative value. In NVIDIA's
design the lasers that feed it sit in a temperature-controlled module
\cite{nvels}, so it is the ring's drift, not the laser's, that the link
must track. Once a ring drifts past the largest
offset the receiver tolerates, the receiver loses lock on it. Parts in
the field stabilize the ring by reactive dither locking: the controller
varies the ring's setting slightly and reads how the output responds.
That response shows which way the resonance has moved, so the correction
comes only after an offset has appeared. A ring reacquires lock in roughly 200~\textmu s \cite{ref28}, and its link carries nothing until it
does. Integrated thermal stabilization of individual microring modulators has
also been demonstrated in hardware \cite{ref21}, model-predictive and
feedforward control of microring modulators has been designed in
simulation \cite{nr_cherian24}, and locking driven by the bit error rate
now tracks 71.4~nm/s with no measurable error-rate penalty under a
40~\degC{} disturbance \cite{nr_shi26}. Closest to ThermaLead, Yoon et
al.\ show that the compute bursts of MoE training heat a wafer-scale
optical interconnect by several kelvin within milliseconds and stall its
rings unless the heater is pre-driven from GPU kernel launches or the
tracking loop runs much faster \cite{nr_yoon26}. ThermaLead instead
forecasts the drift from temperature telemetry.

Lumentum has demonstrated two external laser modules delivering 16
channels on a 200~GHz grid at about 24~dBm per channel
\cite{lumentum26}, and Sirius-class tunable lasers switch within 912~ps
\cite{ref15}. However, the open issue for the laser modules is
reliability: how many spare lasers a pool must hold. In \cite{ref12}, it
was showed that an integrated laser, which cannot be cooled and runs at
the 70~\degC{} maximum chip temperature, needs a backup laser to meet a
1\% annual failure-rate target, whereas an external source can be cooled
by a thermoelectric cooler \cite{ref12}. Pooling needs far fewer spares
than a backup per laser: tunable-laser sparing restores a 32-wavelength
link with only 4--8 shared spares under moderate load
\cite{ref9}. Neither analysis models correlated laser failures, so both
correspond to the zero-correlation case, $\rho \to 0$, to which the
coverage model of Section~\ref{sect:lr} reduces. Studies of distributed
storage show the cost of that omission. In Google's storage fleet,
failing to model correlated node failures typically overestimated
availability by at least two orders of magnitude \cite{ford10}, and
random replica placement is almost guaranteed to lose data when a
cluster-wide power outage takes many nodes down at once
\cite{cidon13}. We found no published vendor pool-sizing rule with a
coverage guarantee.

\paragraph{Formal tools} Coverage bounds, the MULTIFIT
packing ratios of Section~\ref{sect:ff}, one-step-ahead prediction, and
threshold protection are the four formal tools this paper builds on. All
four have a long record in optical WDM survivability and scheduling
\cite{ref5, ref6, ref7, ref8, ref9}. That work targeted fiber cuts at
gigabit rates. The AI data center adds correlated thermal stress on
shared laser pools and on microrings running at 200~Gb/s per lane, and
each component here is designed anew for that setting.

\section{Laser Pooling: The Correlation Ceiling}
\label{sect:ceil}

Pooling spares works like insurance, and it fails for the same reason.
Ten thousand houses in different towns burn independently, so a small
reserve covers the year's claims. Ten thousand houses on one floodplain
produce every claim in a single flood, and insuring more houses there
adds exposure without diversifying it. Co-packaged lasers that share cooling and power resemble the floodplain: one shared cause can bring a burst of failures that
overwhelms spares sized for scattered ones.

\subsection{Mathematical Model}

\paragraph*{Definitions} We need to know whether the spares in a pool
cover the lasers that fail in a window. Consider a pool of $N$ lasers
whose failures are exchangeable, meaning that no laser is special: any
set of $k$ lasers is as likely to be the failing set as any other.
Failures are counted over a fixed window of time: $F$ is the number that
fail in one window, and $\mathcal{F} = F/N$ is the failure fraction. The
window is the restock interval, the time a pool runs before its failed
modules are replaced and its spares refilled, and every coverage and
availability figure in this paper is a per-window quantity. One laser's
probability of failing during one window is its marginal failure
probability $\lambda$, which later sections also call its per-window
failure rate, so $\lambda = 0.05$ means a one-in-twenty chance per laser
per window. Laser reliability is usually quoted as a FIT rate, failures per
$10^{9}$ device hours, and a laser at $\varphi$ FIT with a window of
$t_{\mathrm{w}}$ hours has $\lambda = 1 - \exp(-\varphi\,t_{\mathrm{w}}
\times 10^{-9})$. The failure correlation $\rho \in (0,1)$, the pairwise
(intraclass) correlation between failures, measures how strongly
failures cluster. As $\rho \to 0$ the failures become independent. A larger $\rho$ means that a greater share
of failures traces to one cause acting on many lasers at once. Against
those failures the pool holds $s$ spare lasers, giving a spare ratio
$\sigma = s/N$. Deployments cap the ratio, and we write the budget cap
as $\sigma_{\max}$; the cap assumed throughout is the $2\times$ pool, in
which spares equal active lasers. The severity of the shared cause
varies from window to window, and we model it with the beta distribution
whose mean is $\lambda$ and whose intraclass correlation is $\rho$. Its
shape parameters are $a = \lambda(1-\rho)/\rho$ and $b =
(1-\lambda)(1-\rho)/\rho$. Laser availability, which we also call coverage, is the probability that
the spares cover every failure in a window, written $A_N(\sigma)$. The rest of the paper states it as a percentage in the industry's notation, such as 99.999\%, and figures plot
     it on a logarithmic scale of $1 - A_N$ labelled the same way.
Table~\ref{tab:notation} lists the notation used throughout the paper.

\begin{table}[!t]
\centering
\footnotesize
\setlength{\tabcolsep}{3.5pt}
\begin{tabular}{@{}ll@{}}
\toprule
Symbol & Meaning \\
\midrule
$N$ & total pool size (active plus spare lasers) \\
$n$ & active channels the pool must feed \\
$s$; $\sigma$ & spare count; spare ratio $\sigma = s/N$ \\
$\lambda$ & per-laser failure probability in a window \\
$\varphi$; $t_{\mathrm{w}}$ & laser FIT rate; window length (hours) \\
$\rho$ & pairwise (intraclass) failure correlation \\
$a$, $b$ & Beta parameters of the latent failure level \\
$F$; $\mathcal{F}$ & failures in a window; failure fraction $F/N$ \\
$\Theta$ & latent common-cause failure level, $\Theta \sim \Beta(a,b)$ \\
$A_N(\sigma)$ & availability of an $N$-laser pool at spare ratio $\sigma$ \\
$\Astar$ & availability ceiling $\Fbeta(\sigma;a,b)$ \\
$A_{\mathrm{tgt}}$ & coverage target asked of the sizing algorithm \\
$\sigma_{\max}$ & budget cap on the spare ratio ($2\times$ pool: $1/2$) \\
$A_{\mathrm{crit}}$ & ceiling at the cap, $\Fbeta(\sigma_{\max};a,b)$ \\
$s^{*}$; $s_{5}$ & spares LightReserve returns; spares for 99.999\% \\
$M$; $L$ & modules per pool; lasers per module \\
$\rho_{\mathrm{within}}$, $\rho_{\mathrm{across}}$ & correlation within / across modules \\
$\rhobar$ & pool-average pairwise correlation \\
$\tau$ & threshold: lock (ThermaLead), risk (TriageGuard) \\
$B$ & pre-spare budget (TriageGuard) \\
$\delta$; $m$ & switching delay; circuits per configuration \\
$\sigma_x$; $R$ & ring-offset std.\ dev.; lag-1 autocorrelation \\
\bottomrule
\end{tabular}
\vspace{1.6ex}
\caption{Notation used throughout the paper.}
\label{tab:notation}
\end{table}

\begin{theorem}[Correlation ceiling]   For the system described above,
  all of the following hold:

  \begin{enumerate}

    \item $\mathcal{F}$ converges almost surely to a random level
      $\Theta \sim \Beta(a,b)$, and $\Var(\mathcal{F}) \to
      \lambda(1-\lambda)\rho > 0$.

 \item Availability, $A_N(\sigma) = \Pr[F \le \sigma N] \to
   \Astar(\sigma;\lambda,\rho) = \Fbeta(\sigma;a,b)$.

   That is, as a laser pool operating at a fixed value of $\sigma$ grows,
its availability approaches a limit below one that does not depend on
$N$.

 \item Given a budget cap on the spare ratio, $\sigma \le \sigma_{\max} <
   1$, the availability a pool can approach as it grows is at most
   $\Astar(\sigma_{\max};\lambda,\rho) < 1$, so for any target above it,
   every large enough pool held to that cap falls short of the target.

   \item For every $\sigma > \lambda$, $\Astar \to 1$ as $\rho \to 0$, so the
     ceiling of a $2\times$ pool reaches any target once $\rho$ is low
     enough.

  \end{enumerate}
     \label{thm:ceil}  
\end{theorem}

Under independent failures the variance of the failure fraction shrinks
as $1/N$, so a bigger pool is a safer pool. Part~(1) says that under
correlation the same variance flattens onto a floor, and past a few
hundred lasers the extra hardware buys almost nothing. 

\begin{proof}
Everything turns on one object, the latent common-cause level $\Theta$
that de Finetti's theorem \cite{aldous85} guarantees behind any
exchangeable failure
process that can be extended to an arbitrarily large pool. By de Finetti, an infinite exchangeable binary sequence is a
mixture of i.i.d.\ Bernoulli sequences: there is a latent $\Theta$ with
$X_i \mid \Theta$ i.i.d.\ Bernoulli$(\Theta)$, where $X_i$ indicates the
failure of laser $i$. Read operationally, this says that if no laser is special and the pool
could be extended indefinitely, it must behave as though a hidden
severity level $\Theta$ were drawn first for the window, after which
each laser fails independently at that level. Correlation is then what
happens when $\Theta$ varies between windows, and the beta-binomial
model assumes that a single scalar, $\rho$, is enough to describe how
much it varies. The beta-binomial model is the case $\Theta
\sim \Beta(a,b)$, with the parameters chosen so that $\Ebb[\Theta] =
\lambda$ and $\Var(\Theta) = \lambda(1-\lambda)\rho$, which makes the
pairwise correlation of the $X_i$ equal to
$\Var[\Theta]/[\lambda(1-\lambda)] = \rho$. Each part below is a
statement about that latent level.

\emph{Part~(1).} Conditional on $\Theta$, the failure count $F = \sum_i X_i$
is $\mathrm{Binomial}(N,\Theta)$, so the strong law of large numbers gives
$F/N \to \Theta$ almost surely, and unconditionally the limit is
$\Theta$. The law of total variance then gives
\begin{align*}
\Var(F/N) &= \frac{\Ebb[\Theta(1-\Theta)]}{N} + \Var[\Theta]\\
  &= \frac{\lambda(1-\lambda)[1+(N-1)\rho]}{N},
\end{align*}
whose limit is $\lambda(1-\lambda)\rho$ ($= 0.0071$ at the operating point of
Section~\ref{sect:analysis}).

\emph{Part~(2).} Coverage at spare ratio $\sigma$ holds if and only if
$F/N \le \sigma$, so, since $F/N \to \Theta$ almost surely and $\Fbeta$
is continuous, $A_N(\sigma) \to \Pr[\Theta \le \sigma] =
\Fbeta(\sigma;a,b)$. That limit stays below one for $\sigma < 1$,
because $\Beta(a,b)$ puts positive mass on $(\sigma,1]$.

\emph{Part~(3).} Since $A_N(\sigma)$ is nondecreasing in $\sigma$, a
pool held to $\sigma \le \sigma_{\max}$ has $A_N(\sigma) \le
A_N(\sigma_{\max})$, which converges to $\Fbeta(\sigma_{\max};a,b) < 1$
by Part~(2); the cap bounds the limit, not every finite $N$.

\emph{Part~(4).} As $\rho \to 0$, $a + b = (1-\rho)/\rho$ grows without
bound and $\Var(\Theta) = \lambda(1-\lambda)\rho \to 0$, so $\Theta$
concentrates at its mean $\lambda < \sigma$ and $\Astar \to 1$.
\end{proof}

Parts~(2) and (3) describe the limit, and a finite pool can sit slightly
above it: at the stress point, a large pool at the four-laser reserve's
spare ratio of $1/9$ approaches 85.1\%, below the coverage of the
32-channel reserve itself (Section~\ref{sect:lr}). At the $2\times$ cap
and $\rho = 0.15$, every pool size we evaluated approaches the ceiling
from below, so there the ceiling of a $2\times$ pool is a true upper
limit. Part~(4) states only the limit. Concentration alone does not make $\Astar$
monotone in $\rho$ at every $\sigma$: for a spare ratio only slightly
above $\lambda$, a more correlated pool can hold more of its mass below
$\sigma$, as the left panel of Fig.~\ref{fig:ceiling} shows below a
spare ratio of about 14\%. At $\sigma = 1/2$ we verified numerically
that $\partial\Astar/\partial\rho < 0$ over $\rho \in (0,1)$ for every
$\lambda \le 0.3$, so at the $2\times$ cap the ceiling $\Astar$ strictly increases as $\rho$ falls.

\begin{figure}[!t]
\centering
\includegraphics[width=\linewidth]{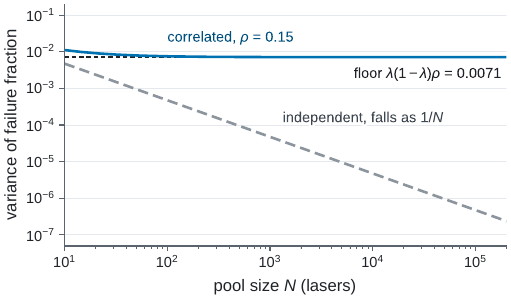}
\caption{Why correlated failures do not average out: the variance of
the failure fraction $\mathcal{F}$ against pool size $N$, on
logarithmic axes. With independent failures (falling line) the variance
shrinks as $1/N$. With correlation ($\rho =
0.15$, upper curve) it flattens at the floor
$\lambda(1-\lambda)\rho = 0.0071$ [derived].}
\label{fig:selfavg}
\end{figure}

\subsection{Analysis}
\label{sect:analysis}

The theorem supplies no value for $\lambda$ or $\rho$, and every number
below moves with both. The
analysis takes a stress operating point of $\lambda = 0.05$ and $\rho =
0.15$. When the
per-laser failure probability is small, $\rho$ is the common-cause beta factor of reliability engineering: the share of failures that trace
back to a single shared cause. Other industries have estimated it. IEC~61508-6 assigns values from
0.005 to 0.05 for programmable electronics and from 0.01 to 0.10 for
field equipment, according to the defenses in place \cite{iec61508-6},
and nuclear practice uses 0.05 to 0.10 for a redundant pair
\cite{nureg5485, hark16}. Where the units share one die, IEC~61508-2
starts from 0.33 and accepts the redundancy as adequately separated only
once the estimate is shown to be below 0.25 \cite{iec61508-2}.

Lasers in one module have little physical separation, since they share
its enclosure and temperature control and often one power supply, so a
pool built from a few such modules sits away from the well-separated
end of that range. We therefore take $\rho = 0.15$,
between the 0.10 upper end of the IEC~61508-6 and nuclear ranges and the
0.25 on-die limit, as an estimate for a well-engineered
package, a pool-wide value that Section~\ref{sect:hier} splits into
within-module and across-module parts. It is an assumption, not a measurement: no field
data for these laser form factors pins it down yet, and
Section~\ref{sect:measure} describes the experiment that would. The
choice is pessimistic for availability and favorable to our argument,
since a lower $\rho$
would raise the ceiling of Theorem~\ref{thm:ceil} and so weaken the case
for decorrelating. We also report the full $\rho$-sweep, including a plausible band of
$\rho \in [0.08, 0.35]$ for the cost results.

The failure probability needs the same care, because the window turns a
FIT rate into a probability. The OIF co-packaging framework sets a reliability
target of 50~FIT for the laser source of each 3.2~Tb/s engine over a
six-year service life \cite{oifcpo}. The target covers a whole source,
and a source can hold several lasers, as NVIDIA's eight-laser module
does, so applying 50~FIT to each laser is a pessimistic reading of it. An average optical transceiver runs at about 200 to 300~FIT, and low-cost lasers recently deployed in data centers failed at
ten times their target rate \cite{herrick}. Against those figures,
$\lambda = 0.05$ is a stress setting: it is what a laser at about 1,000~FIT accumulates when it is never replaced over a six-year life.
For co-packaged optics, Broadcom reports one million 400G-equivalent
port-hours without a link flap in Meta's high-temperature lab
\cite{bcm25cpo}, which caps the per-port flap rate near 3,000~FIT at
95\% confidence, above both 50~FIT and the stress setting.
Table~\ref{tab:regime} evaluates the same pool at rates nearer the
targets. For the 32-channel pool at $\rho = 0.15$, 99.999\% stops
fitting inside the $2\times$ cap once $\lambda$ exceeds about $2.4
\times 10^{-4}$ per window. That happens when the FIT rate times the restock interval passes about $2.4 \times 10^{5}$ FIT-hours, after
about 200 days at 50~FIT and about 40 days at 250~FIT.

\begin{table}[!t]
\centering
\footnotesize
\setlength{\tabcolsep}{3.2pt}
\begin{tabular}{@{}rrrrrr@{}}
\toprule
$\varphi$ (FIT) & Window & $\lambda$ & $A_{\mathrm{crit}}$ (\%) & $s_{5}$, $\rho = 0.15$ & $s_{5}$, indep. \\
\midrule
50 & 90 days & $1.1 \times 10^{-4}$ & 99.9996 & 25 & 1 \\
50 & 1 year & $4.4 \times 10^{-4}$ & 99.9985 & 39 & 2 \\
250 & 1 year & $2.2 \times 10^{-3}$ & 99.9922 & 61 & 3 \\
500 & 1 year & $4.4 \times 10^{-3}$ & 99.9839 & 73 & 4 \\
976 & 6 years & 0.05 & 99.6697 & 156 & 10 \\
\bottomrule
\end{tabular}
\vspace{1.6ex}
\caption{Where the ceiling binds, for the 32-channel pool. Each row
turns a laser FIT rate $\varphi$ and a restock interval $t_{\mathrm{w}}$
into the per-window probability $\lambda$. $A_{\mathrm{crit}}$ is the
large-pool ceiling of a $2\times$ pool (Corollary~\ref{cor:crit}), and
the 32-channel pool reaches slightly less. $s_{5}$ is that pool's spare
count for 99.999\% coverage, correlated and independent; the $2\times$
cap allows 32, which sets the $2.4 \times 10^{-4}$ threshold. The
250~FIT row sits in the range of an average optical transceiver, the
500~FIT row is ten times the OIF target, and the last row is the stress
operating point used throughout [derived].}
\label{tab:regime}
\end{table}

Fig.~\ref{fig:selfavg} shows Part~(1) at this operating point: the
failure fraction keeps a standard deviation of 0.084 however large the
pool grows.

\begin{figure*}[!t]
\centering
\includegraphics[width=\linewidth]{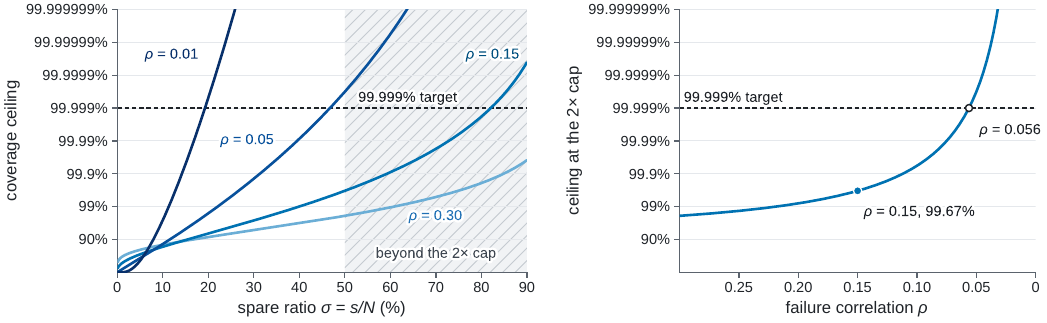}
\caption{The correlation ceiling. Left: the coverage ceiling against the spare ratio $\sigma$, one curve per correlation $\rho$. The dashed line is the 99.999\% target and the hatched band marks spare ratios past the $2\times$-pool budget cap. At $\rho = 0.15$ the curve reaches the cap below the target. Right: the ceiling a fixed $2\times$ pool attains as $\rho$ falls (correlation decreases to the right); the filled dot marks the $\rho = 0.15$ operating point at 99.67\%, and the open dot the correlation, about 0.056, at which the pool reaches 99.999\%. Spares move along one curve, and only decorrelation moves across curves to 99.999\% [derived].}
\label{fig:ceiling}
\end{figure*}

In the left panel of Fig.~\ref{fig:ceiling}, 99.999\% availability at
this operating point requires a spare ratio of $\sigma = 0.821$.
Since $\sigma = s/N$, the total pool must then be $1/(1-\sigma) \approx
5.6\times$ the active laser count, which no deployment would carry; that is the large-pool limit, and
Section~\ref{sect:lr} gives the finite-pool answer in whole lasers.

\begin{corollary}[Critical availability]
Given a budget cap $\sigma_{\max}$, the value $A_{\mathrm{crit}} =
\Fbeta(\sigma_{\max};a,b)$ acts as a threshold for large pools: a target
below $A_{\mathrm{crit}}$ is reachable with a spare ratio inside the cap
once the pool is large enough, while a target at or above it cannot be
reached by any ratio inside the cap, and the pool multiple
$1/(1-\sigma)$ it would demand grows without bound as the target
approaches one. For a $2\times$ pool at the stress rate, $A_{\mathrm{crit}} =
99.670\%$ at $\rho = 0.15$, rising above 99.999\% once $\rho$ falls
below about 0.056.
\label{cor:crit}
\end{corollary}

At this operating point no spare ratio inside the cap reaches 99.999\%, and a low enough correlation does, since the ceiling of a
$2\times$ pool tends to one as $\rho$ falls (Part~(4) of
Theorem~\ref{thm:ceil}). Section~\ref{sect:split} prices decorrelation, and the right panel of
Fig.~\ref{fig:ceiling} shows the rise for a fixed $2\times$ pool.

Fast recovery offers a partial second route toward 99.999\%, because
it shortens an exhaustion that coverage fails to prevent. Coverage ignores how long an exhaustion
lasts, and \emph{effective availability} also counts how quickly it is
repaired. Take a pool sized for 99.9\% coverage, which is 49 spares
on the 32-channel calibration point of Section~\ref{sect:lr}.  TriageGuard (Section~\ref{sect:tg})
keeps hot spares pre-wired to the modules most likely to fail, so a
failure on such a module switches to its hot spare in milliseconds
instead of lasting the rest of the window, and those hot spares sit
outside the shared pool, so they are still there when the pool runs out.
Only failures on the pre-spared modules are shortened, however. Under
the exchangeable failures of Theorem~\ref{thm:ceil} every module carries
the same risk, so hot spares on 4 of the 18 modules see on average
$4/18$, about 22\%, of the failures in an exhausted window, and a risk
ranking raises that share only as far as module risks differ and are
predicted (Section~\ref{sect:tg}). Fast recovery therefore narrows the
gap to 99.999\% by an amount we have not computed, and replacements
must also keep arriving to restock the pool.

The pairing of coverage and fast recovery has two further limits. First,
those 49 spares make a $2.5\times$ pool before the hot spares are
counted. That is less than a third of the 156 spares the
coverage-only route needs, but the pairing is not shown to reach 99.999\%, and even its starting pool sits above the $2\times$ budget cap,
so fast recovery cannot bring the target inside the cap. Only
decorrelation, priced in Section~\ref{sect:split}, fits inside the cap. Second,
LightReserve counts spares in lasers while TriageGuard hands out hot
spares a module at a time, so combining them in a deployment requires
fixing the modules-of-lasers shape of Section~\ref{sect:hier} first. 

\begin{figure*}[!t]
\centering
\includegraphics[width=\linewidth]{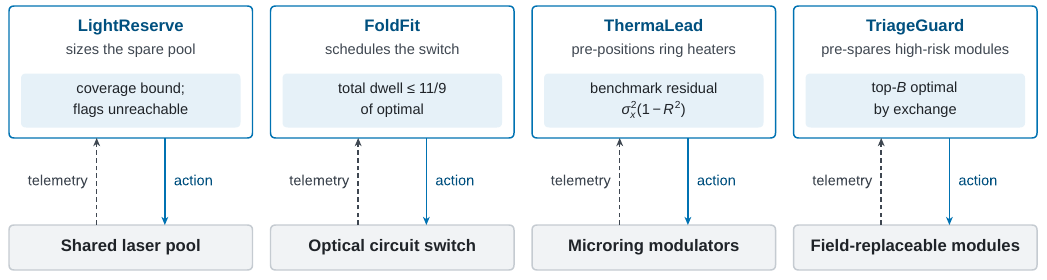}
\caption{The \textsf{Halley} control plane, with one component per data-plane layer: LightReserve over the shared laser pool, FoldFit over the OCS, ThermaLead over the microrings, and TriageGuard over the replaceable modules. Dashed arrows send each layer's telemetry up into a prediction, and solid arrows send the resulting action back down. The shaded band in each component names its formal guarantee in the
idealized form of Table~\ref{tab:components}, or for ThermaLead the
benchmark its forecast is measured against [schematic].}
\label{fig:control}
\end{figure*}

\section{Proposed Control Plane Design}
\label{sect:cpfrwk}

This section presents the details of the proposed \textsf{Halley} framework.
Four components, one per layer, sit above the optical data plane, as
shown in Fig.~\ref{fig:control}. Table~\ref{tab:components} lists the
decision each one makes, its guarantee or, for ThermaLead, its
benchmark, and its reactive fallback.  All four follow one rule: each
acts on a prediction of its layer's next state, and when its forecast is
stale or unsure, it steps aside and the system runs exactly the deployed
policy it replaces. A confident but wrong forecast has no such floor.

\begin{table*}[!t]
\centering
\footnotesize
\setlength{\tabcolsep}{4pt}
\begin{tabular}{@{}lllll@{}}
\toprule
Component & Data-plane layer & Decision it makes & Guarantee or benchmark & Reactive fallback \\
\midrule
LightReserve & shared laser pool & spare count $s^{*}$ & coverage bound; flags unreachable & static $n{+}k$ reserve \\
FoldFit & OCS & configurations and dwell times & total dwell $\le 11/9$ of optimal & BvN or greedy decomposition \\
ThermaLead & microring modulators & heater pre-position each step & benchmark residual $\sigma_x^2(1-R^2)$ & reactive dither locking \\
TriageGuard & field-replaceable modules & which modules get hot spares & top-$B$ optimal by exchange & uniform sparing from the pool \\
\bottomrule
\end{tabular}
\vspace{1.6ex}
\caption{The four control-plane components. 
LightReserve, FoldFit, and TriageGuard carry a formal guarantee, shown
here in idealized form; ThermaLead's entry is a benchmark. FoldFit's $11/9$ holds under the three
conditions of Section~\ref{sect:ff} and leaves out switching delay. ThermaLead's benchmark is the residual variance of the
minimum-mean-square-error forecast from the previous sample for
first-order autoregressive (AR(1)) drift. Its moving-average forecast is
cheaper, and on the stress trace of Section~\ref{sect:tl} it does not
reach this figure. TriageGuard's top-$B$ optimality is conditional on the
risk estimate it is given.}
\label{tab:components}
\end{table*}

\begin{algorithm}[!t]
  \caption{LightReserve pool sizing}
  \label{alg:lr}
\begin{algorithmic}[1]
\State \textbf{input:} active channels $n$, failure rate $\lambda$, correlation $\rho$, target $A_{\mathrm{tgt}}$
\State model total pool $N = n + s$; spares are themselves failure-prone; coverage holds iff $F \le s$
\State $A(s) \gets \Pr[F \le s]$,\quad $F \sim \mathrm{Binomial}(N,\lambda)$ or $\mathrm{Beta\text{-}Binomial}(N,a,b)$
\State $s^{*} \gets \min\{\,s : A(s) \ge A_{\mathrm{tgt}}\,\}$ \Comment{$A(s)$ increasing; binary search}
\If{$s^{*}/(n+s^{*}) > \sigma_{\max}$}
    \State report \textsc{unreachable} and emit the $\rho'$ that brings $A_{\mathrm{tgt}}$ under the ceiling
\EndIf
\State \textbf{output:} $s^{*}$ spares, or a decorrelation target when the ceiling forbids the goal
\end{algorithmic}
\end{algorithm}

\begin{figure}[!t]
\centering
\includegraphics[width=\linewidth]{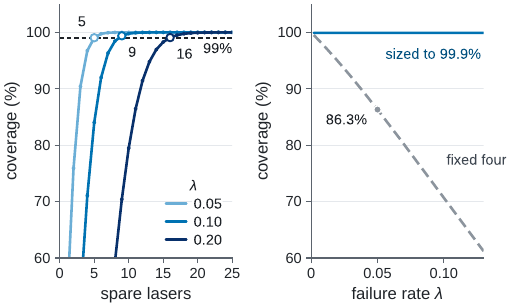}
\caption{LightReserve dimensioning at the 32-channel calibration point. Left: coverage against spare count, one curve per failure rate, with failures assumed independent; the spare count for 99\% coverage varies by $3\times$ across these stress rates. Right: at $\rho = 0.15$, coverage against failure rate for the fixed four-laser reserve and for a pool that LightReserve sizes to a 99.9\% target at each rate [derived].}
\label{fig:lightreserve}
\end{figure}

\subsection{LightReserve: Sizing the Laser Pool}
\label{sect:lr}
LightReserve decides how many spares a shared laser pool carries.
Operators fix the number once, as a static $n{+}k$ reserve, and hold it
whatever the failure rate turns out to be. LightReserve computes it from
the failure rate and from how strongly the failures are correlated, and
returns it with a provable coverage guarantee for those inputs.

Algorithm~\ref{alg:lr} models the total pool $N = n + s$ and not the
active set alone, because the spares are themselves failure-prone. If the spare ratio at $s^{*}$, the smallest count whose coverage meets
$A_{\mathrm{tgt}}$, exceeds the budget cap $\sigma_{\max}$, the algorithm
reports
\textsc{unreachable} and, in place of a spare count, emits the
correlation $\rho'$ that would bring the target under the ceiling, which
turns an impossible sparing problem into a concrete decorrelation
target.

Fig.~\ref{fig:lightreserve} gives numerical results for the calibration
point of Section~\ref{sect:odp}. With independent failures (left panel), LightReserve returns
the classical small spare counts: 5 spares for 99\% coverage at $\lambda
= 0.05$, 9 at $0.10$, and 16 at $0.20$. As $\rho \to 0$ the method
reduces to that known result.

At $\lambda = 0.05$ and
$\rho = 0.15$ the four-laser reserve gives $A(4) = 86.3\%$, and 99\%
coverage needs 23 spare lasers where independence needed 5. Reaching
99.9\% takes 49, and 99.999\% takes 156, a $5.9\times$ pool. The
algorithm flags that pool as \textsc{unreachable} and returns a
correlation target in its place: 99.999\% within a $2\times$ pool
of 32 channels needs $\rho \le 0.046$, a little below the large-pool
value of about 0.056 in Corollary~\ref{cor:crit}. The flag concerns coverage only; fast recovery
(Section~\ref{sect:analysis}) narrows the effective-availability gap,
without a computed guarantee and from a pool still above the cap.

The right panel sweeps the failure rate: the fixed reserve's coverage
falls steeply as the rate grows, while a pool that LightReserve sizes to
each rate holds its 99.9\% target, though above about $\lambda = 0.02$
that sizing needs more than the 32 spares the $2\times$ cap allows. For
99.999\% coverage, which the panel does not plot, the correlation
limits the pool more than the rate does: at $\rho = 0.15$, even $\lambda
= 0.005$, a tenth of the stress rate, still needs 76 spares, more than
twice the 32 the cap allows. Only
a rate below about $2.4 \times 10^{-4}$ brings 99.999\% inside the cap
(Table~\ref{tab:regime}), whereas lowering $\rho$ cuts the count at
every rate.

\begin{algorithm}[!t]
  \caption{FoldFit (binary-searched first-fit-decreasing, FFD, demand decomposition)}
  \label{alg:foldfit}
\begin{algorithmic}[1]
\State \textbf{input:} demand matrix $D$, channel capacity $C$, channel count $m$, MoE block structure
\State jobs $\gets$ nonzero demands of $D$ (in slots at capacity $C$), each tagged with its (row, col) locality
\State $T_{\mathrm{lb}} \gets$ the largest of $\max D$, $\sum D/m$, and the largest row or column sum of $D$
\State $T_{\mathrm{hi}} \gets \max\{T_{\mathrm{lb}},\ \sum D/m + \max D\}$
\State binary-search the deadline $T \in [T_{\mathrm{lb}}, T_{\mathrm{hi}}]$, packing jobs by FFD at each candidate $T$
\State $T^{*} \gets$ the feasible $T$ on which the search ends \Comment{after $q$ steps, $T^{*} \le (13/11 + 2^{-q})\,\mathrm{OPT}$}
\State break ties to keep expert-parallel blocks together, so stable configurations are reused
\State \textbf{output:} configurations $+$ dwell times, makespan $T^{*} \le (11/9)\,\mathrm{OPT}$ for $q \ge 5$, under the conditions stated in the text
\end{algorithmic}
\end{algorithm}

\begin{figure*}[!t]
\centering
\includegraphics[width=\linewidth]{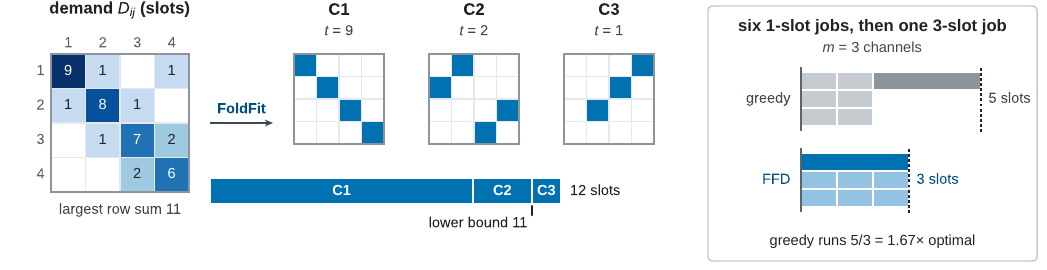}
\caption{FoldFit worked example. Left: the demand matrix of a MoE collective,
each cell a traffic demand measured in slots. Middle: the schedule
FoldFit folds it into, where each filled cell is an active circuit and
$t$ gives each configuration's dwell time. Right: the canonical
worst case for greedy list scheduling, which serves demands in arrival
order [derived].}
\label{fig:foldfit}
\end{figure*}

\subsection{FoldFit: Scheduling the Circuit Switch}
\label{sect:ff}

FoldFit schedules the OCS that joins the photonic engines to one
another. The switch holds only a limited set of connections at a time, so a
communication phase is served as a sequence of configurations, each held
for a while before the switch is reconfigured. A good sequence is short,
because the collective waits while each configuration is held, and
stable, because every reconfiguration costs switching time. FoldFit
chooses the sequence by recasting the problem as packing: each traffic
demand becomes a job, and the jobs are packed onto the parallel circuits
the switch can hold at once.

 The fabric is a single-stage OCS with crossbar semantics. At any
moment it holds one \emph{configuration}, a set of port-disjoint
circuits, meaning that no two of them share a switch port. Traffic
between two engines flows only while a circuit connects them, so serving
a workload means choosing a sequence of configurations and a \emph{dwell
time} for each. Those dwells sum to the makespan of
Section~\ref{sect:odp}. Each change of configuration also costs a fixed
switching delay $\delta$, which the makespan as defined here leaves
out. The input is the demand matrix $D$, with entry
$D_{ij}$ the traffic engine $i$ must send to engine $j$, expressed in
time slots at the per-channel capacity $C$. 

The classical way to choose that sequence is the BvN decomposition
\cite{ref16}, which pads the demand matrix until every row and column
sums to the busiest port's load and writes it as a weighted sum of
permutation matrices, each a configuration whose weight is its dwell.
The dwells then meet the busiest-port bound, but over as many as
$(P-1)^2+1$ configurations for $P$ ports, most with tiny dwells, each
costing a reconfiguration delay that nothing in the decomposition
bounds.

FoldFit replaces the decomposition with a packing, a problem that has
carried a worst-case makespan bound since the 1970s. FoldFit treats each nonzero demand as
an indivisible job of length $D_{ij}$ slots and the fabric as $m$
identical service channels, where $m$ is the number of port-disjoint
circuits one configuration can sustain. These channels are the parallel
machines of the packing problem, not the wavelength channels of the
laser pool. The target is $\mathrm{OPT}$, the shortest schedule that serves all
jobs on $m$ parallel channels without two circuits sharing a port at the
same time. Dropping the port condition leaves the classical makespan
problem, whose optimum is no longer.

Three lower bounds on that optimum are immediate. No schedule beats the
largest single demand; none beats the average load $\sum_{ij} D_{ij}/m$;
and none beats the busiest port's total traffic, the largest row or
column sum of $D$, because a port carries one circuit at a time.
Starting from those bounds, Algorithm~\ref{alg:foldfit} binary-searches
the deadline $T$, a candidate makespan, running FFD at each candidate $T$. FFD sorts the jobs from longest to shortest
and places each in turn on the first channel whose load would stay
within $T$. If every job fits, $T$ is feasible. The deadline $T^{*}$ on which the search ends fixes the assignment of jobs to
channels.

Binary-searched FFD is the MULTIFIT algorithm of multiprocessor
scheduling, here with its search started at $T_{\mathrm{lb}}$, and it
inherits the MULTIFIT theorem. On identical channels,
$q$ steps of the search give $T^{*} \le (1.22 + 2^{-q})\,\mathrm{OPT}$
\cite{ref17}, and Yue tightened the constant to the exact ratio, $T^{*}
\le (13/11 + 2^{-q})\,\mathrm{OPT}$ \cite{ref26}. Both bounds are stated against the classical optimum, which is no longer
than $\mathrm{OPT}$. Every deadline at which FFD fails lies below
$13/11$ of the classical optimum, the search starts at $T_{\mathrm{lb}}
\le \mathrm{OPT}$, and its upper end, $\max\{T_{\mathrm{lb}}, \sum D/m +
\max D\}$, keeps the search interval no wider than the classical
optimum, so both ratios hold against $\mathrm{OPT}$. They need not hold
against the classical optimum itself, which the busiest-port bound can
exceed several times over. With five or more
steps Yue's guarantee sits below $11/9$, the figure we quote throughout.
Under the three conditions below, the schedule is then never more than
22\% longer than the optimum however the demand matrix moves, and the
exact bound narrows that margin toward 18\% as $q$ grows.

MULTIFIT is a result about $m$ identical machines, and it applies to a
switch under three conditions. First, the channels have to be
interchangeable, which holds while every circuit runs at the same
per-channel capacity $C$. Second, each demand has to be packed as one indivisible job on one
channel. Third, FFD's assignment of jobs to channels has to be
realizable as a sequence of port-disjoint configurations.

The classical setting does not supply the third condition, because two
jobs sitting on different channels may share a switch port, and no
configuration can run both at once. We call the step that turns the
assignment into port-disjoint configurations the alignment. The
busiest-port bound is necessary for an alignment to exist at all, which
is why the binary search starts no lower than it.  When a demand may be split at a
configuration boundary, which is the relaxation BvN works in, the alignment is exact and the $11/9$ guarantee carries over untouched. When demands
are strictly unsplittable, the alignment can leave idle gaps that the
$11/9$ bound does not account for, and the worked example below loses
one slot this way. On the workloads swept in Section~\ref{sect:simmeth}, FoldFit's makespan lands within 6\% of the larger of the first two lower
bounds above on every instance, and within about 1\% on average from 16
engines up. The simulation packs jobs onto channels without the port
condition, so that evidence covers the packing and says nothing about
alignment gaps; Section~\ref{sect:limits} records the unsplittable case
as open.

One refinement adapts the packing to MoE traffic. Expert-parallel demand
is block-structured and drifts slowly as expert popularity shifts. When several
placements tie during FFD, FoldFit keeps jobs from the same
expert-parallel block in the same configuration, so consecutive
scheduling rounds reproduce the same configurations and the switch can
hold them instead of rebuilding them. The simulation of Section~\ref{sect:simmeth} replaces first fit with
this preference and, failing it, the least-loaded channel, so it
evaluates a variant that the $11/9$ proof does not cover. Its baseline, a longest-processing-time (LPT) list schedule that places
one demand in five on a random channel, spreads every block over every
channel. Against it, FoldFit cuts a proxy for reconfigurations, the
number of distinct expert blocks each channel serves, by 45\% at 16
engines. At 8 engines, a single block, the two tie.

Fig.~\ref{fig:foldfit} works through a $4\times4$ MoE demand matrix,
dense on the expert-block diagonal with entries of 9, 8, 7, and 6 slots
and light off the diagonal. The busiest port carries 11 slots, so no
schedule is shorter than 11, and Algorithm~\ref{alg:foldfit}'s search
stops at that bound, $T^{*} = 11$. Turning the packing into port-disjoint configurations costs one more
slot: C1 carries the diagonal entries and dwells 9 slots, C2 the
exchanges inside each two-engine expert block for 2, and C3 the
cross-block residual for 1. The makespan is 12 slots before switching
delay, a ratio of 1.09 to the lower bound; the extra slot is an
alignment loss that the $11/9 \approx 1.22$ guarantee does not cover,
and the ratio still falls inside it. A BvN
decomposition of the same matrix meets the 11-slot bound on dwell by
splitting demands; the decomposition we computed uses five
configurations to FoldFit's three, so it pays the switching delay
$\delta$ five times instead of three.

\paragraph{How this differs from MixNet} MixNet \cite{ref14} exploits
the same expert-parallel locality and reconfigures the switch in advance
from a prediction of the first all-to-all's demand, but gives no
worst-case bound, so on an adversarial shift its makespan is
unconstrained. FoldFit adds a bound that holds however MoE traffic moves
mid-training, under the three conditions above.

\paragraph{How this compares with circuit-switch schedulers} Sunflow
\cite{sunflow} finishes a single coflow within $2\times$ of the optimal time for any
reconfiguration delay $\delta$, and its bound already counts the delay
of every reconfiguration it makes. FoldFit's $11/9$ is tighter but narrower. It bounds the
dwell time $T^{*}$, and a schedule with $k_{\mathrm{r}}$
reconfigurations takes $T^{*} + k_{\mathrm{r}}\delta$ in all, a second
term the bound does not cover. Commercial OCSs typically switch in
10--20~ms \cite{apollo22}, so with the worked example's three
configurations that term stays under 1\% of the dwell only for phases
of 3--6~s or longer, and rescheduling within one MoE layer's dispatch
would need microsecond-class or faster switching \cite{ref15}. Keeping
the shorter of the FoldFit and Sunflow schedules bounds the total by the
smaller of $(11/9)\,\mathrm{OPT} + k_{\mathrm{r}}\delta$ and twice
Sunflow's optimum, in Sunflow's switch model. Solstice \cite{solstice}
minimizes a similar total, dwell time plus $\delta$ per configuration,
by heuristic and without a worst-case guarantee. Eclipse
\cite{eclipse} maximizes the traffic served within a fixed window, a
different objective.

\paragraph{Why the tighter bound matters}  Plain LPT list scheduling carries Graham's $4/3 - 1/(3m)$ ratio
\cite{ref18}. That ratio equals $11/9$ at $m = 3$ and is tighter at $m =
2$, so the quoted $11/9$ improves on it only for $m \ge 4$, while Yue's
exact $13/11$ improves on it for every $m \ge 3$.  The right panel of Fig.~\ref{fig:foldfit} shows the canonical
worst case for list scheduling: six unit demands followed by one
three-slot demand on $m = 3$ channels. A greedy scheduler serving
demands in arrival order spreads the six unit jobs across all three
channels and must then run the long job alone, for a makespan of 5 slots
against an optimum of 3. That is $5/3 \approx 1.67\times$ optimal,
exactly Graham's $2 - 1/m$ bound for list scheduling at $m = 3$
\cite{ref18}. FFD's decreasing order places the long job first and stays
exactly optimal here. Expert popularity shifting mid-training or
mid-serving can produce orderings of exactly this shape.

\subsection{ThermaLead: Steadying the Microring}
\label{sect:tl}

The receiver loses lock once the ring's resonance offset exceeds the
lock threshold, the largest offset it tolerates. Parts in the field use the
reactive dither locking of Section~\ref{sect:odp}, which corrects an offset only after observing it, so the receiver absorbs each step's drift before the correction lands. ThermaLead adds a
feedforward term. It predicts the next-step drift from a short history and
pre-positions the heater one step early, so the receiver sees only the error of that forecast. The heater and its authority
are the same in both schemes; only the excursion, the swing of the
offset at the receiver, changes.

\begin{algorithm}[!t]
  \caption{ThermaLead one-step-ahead microring control}
  \label{alg:tl}
\begin{algorithmic}[1]
\State \textbf{input:} per-ring temperature history, ring slope 70~pm/\degC, lock threshold $\tau$
\State $\hat{x}[t{+}1] \gets$ one-step predictor (8-sample moving average of $dT/dt$)
\State pre-position the heater to cancel $\hat{x}[t{+}1]$
\State residual $r[t{+}1] = x[t{+}1] - \hat{x}[t{+}1]$ \Comment{only what the prediction missed}
\end{algorithmic}
\end{algorithm}

In Algorithm~\ref{alg:tl}, $x[t]$ is the ring's resonance offset at
control step $t$, in picometers via the 70~pm/\degC{} slope. The 8-sample
moving average of line 2 costs an adder and a shift, so it is cheap
enough to sit beside the local servo, the circuit that sets the heater
from the temperature error, which Section~\ref{sect:hwimp} builds in
hardware. The residual $r[t{+}1]$ of line 4 is the part of the drift
that the predictor missed and that is left for the receiver.

ThermaLead's benchmark is the residual of the best one-step predictor
from the previous sample. We model the drift $x[t]$ as
wide-sense stationary over the window the predictor sees. Its variance is $\sigma_x^2$
(this subsection's $\sigma$ is a standard deviation of the resonance
offset, not the spare ratio of Section~\ref{sect:ceil}), and its lag-1
autocorrelation $R$ is the correlation between one step's offset and the
next. By the orthogonality principle of estimation theory, the
minimum-mean-square-error predictor leaves residual
variance $\sigma_e^2 = \sigma_x^2(1 - R^2) < \sigma_x^2$, the best any
predictor can do from the previous sample alone. The expression is exact for AR(1) drift; the evaluation's trace adds a periodic heat cycle to an AR(1) process
with noise bursts, so on that trace it is an approximation. The moving
average of line 2 is a cheaper forecast that extrapolates the recent
slope. It uses eight samples, so the bound above does not cover it, and
the simulation below measures what it achieves.

Treat the offset as Gaussian, a modeling assumption that the trace
below satisfies only approximately. An offset of standard deviation
$\sigma$ then exceeds the lock threshold $\tau$ with probability
$2\bar{\Phi}(\tau/\sigma)$ on any given step, where $\bar{\Phi}$ is the
standard normal upper-tail probability. Shrinking $\sigma$ from
$\sigma_x$ to $\sigma_e$ therefore strictly lowers the lock-loss rate,
and the larger $R$ is, the larger the gain.

The lock-loss argument fixes neither the threshold nor the control step. The
lock-loss rate depends on the threshold only through the ratio
$\tau/\sigma$, so shrinking $\sigma$ lowers the rate wherever a
particular receiver's threshold sits.  The control step is only a unit of counting: the analysis and the
counts below are both per step, and any reduction is a ratio of two
per-step counts, so no particular duration is assumed. The step must be short
against the thermal time constant, so that a one-step-ahead forecast has
something to predict, and long enough for the heater to act.
Section~\ref{sect:hwimp} shows that the heater drive is not the binding
constraint; the heater's own thermal response is.

The evaluation models the thermal environment of training as a periodic
heat cycle, a sinusoid with an 8~\degC{} swing and a period of 120
control steps. The cycle stands for the bulk-synchronous alternation in
which each compute phase heats the die and each communication phase lets
it cool. On top of it sits an AR(1) residual with coefficient 0.8, whose noise
has a standard deviation of 0.55~\degC{} on a random 30\% of steps and
0.12~\degC{} on the rest. The lock threshold is 25~pm. We treat this trace as a stress case.

In simulation the deployed dither lock cancels every offset once it has
observed it, so the receiver sees each step's full drift.  On the stress
trace the dither lock loses lock on 462 of every 2000 control steps and
ThermaLead on 472, averaged over 30 seeds. Paired seed by seed,
ThermaLead loses lock 2\% more often ($p = 0.003$, paired $t$-test) and
does worse on 22 of the 30 seeds. Its residual variance is 1.3\% above
the drift's own, so its forecast removes nothing. The closed form above
explains part of this. The drift's lag-1 autocorrelation is only 0.075,
so even the best linear forecast from the previous sample could remove
less than 1\% of its variance. The trace is not unforecastable, however.
An oracle forecast that knows the trace's sinusoid and noise state, a
bound and not a controller, would cut the count to 315, 32\% below the
dither lock. ThermaLead gains as the drift gets smoother. When noise
bursts hit 10\% of steps instead of 30\%, it loses lock 22\% less often
than the dither lock, and with no bursts 78\% less often. A gate on the forecast's own error removes the stress-trace deficit.
Applied only while its mean absolute error over the last 16 steps is
below the raw drift's, the gated controller loses lock on 452 steps,
2.1\% fewer than the dither lock and better on 23 of the 30 seeds, and
keeps at least 95\% of the forecast's gain on smoother traces. The rule
was fixed before any run; other windows and error measures give 448 to
487, so the result depends on it. On the burst sweep the drift's lag-1
autocorrelation was 0.13 or more wherever ThermaLead helped and 0.075
where it lost, so ring telemetry could tell an operator which case
holds.

ThermaLead is therefore the least proven of the four: its value depends
on how smooth the field's drift is, and Section~\ref{sect:limits} shows its ungated deficit growing under abrupt transients. ThermaLead also serves as a
sensor. The per-ring temperature telemetry it samples every control step
is the newest thermal signal in the package, and TriageGuard's risk
ranking runs on it.

\subsection{TriageGuard: Protecting the Modules}
\label{sect:tg}

A fixed budget of hot spares cannot protect every module, so the budget
should go to the modules most likely to fail. Deployed systems hand out
spares uniformly from the shared pool. TriageGuard
scores each module on its temperature, its age, and its observed
forward-error-correction (FEC) bin, the band of corrected-error counts
the link reports. The result is the module's risk score, the estimate of
how likely it is to fail. It then pre-spares every module above
a risk threshold $\tau$, up to the pre-spare budget $B$, the number of
modules that can hold a hot spare, and routes the rest to the shared
pool.

Protecting the top-$B$ risks is optimal, by an exchange argument. The
expected number of protected failures is additive over the protected
set, a sum of one term per protected module. Swapping a protected
low-risk module for an unprotected higher-risk one never lowers the sum,
and repeating such swaps turns any selection into the top-$B$ selection.
Greedy top-$B$ selection is therefore exactly optimal for any fixed risk
estimate. The argument settles how to spend the budget once the modules
are ranked and says nothing about whether the ranking is right, so
optimality is conditional on the risk estimate, and the freshness of
that estimate dominates the result below.

\begin{algorithm}[!t]
\caption{TriageGuard risk-ranked selective protection}
\label{alg:tg}
\begin{algorithmic}[1]
\State \textbf{input:} risk scores $r_1 \ge \cdots \ge r_M$, budget $B$, threshold $\tau$, switch $t_{\mathrm{sw}}$ (5~ms), miss $t_{\mathrm{miss}}$ (150~ms)
\State pre-spare the top-$B$ modules with $r > \tau$ \Comment{fast path}
\State route the remainder to the LightReserve pool \Comment{dynamic path}
\State on a failure, recovery $= t_{\mathrm{sw}}$ if pre-spared, else $t_{\mathrm{miss}}$
\end{algorithmic}
\end{algorithm}

\begin{figure*}[!t]
\centering
\includegraphics[width=\linewidth]{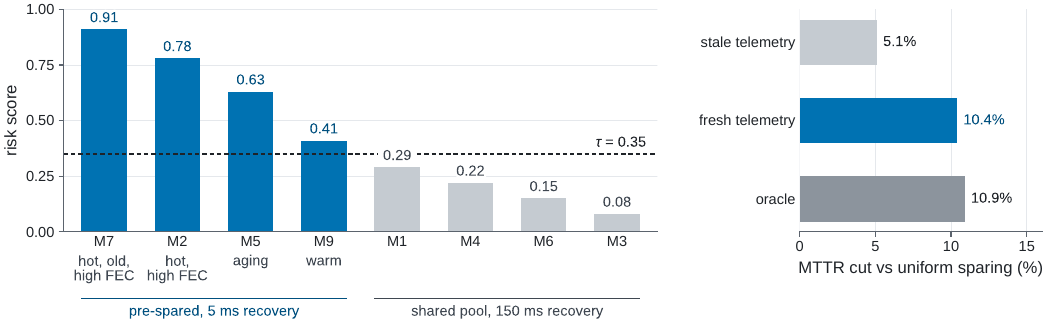}
\caption{TriageGuard worked example. Left: the risk scores of eight of the pool's
modules, with the risk threshold $\tau = 0.35$ dashed; the four modules
above it get pre-allocated hot spares, and the rest fall back to the
shared pool. Right: in simulation, the cut in mean time to recover
against uniform sparing of the same four-module budget, for rankings on
stale and on fresh telemetry and for an oracle ranking that knows each
module's failure risk [composed].}
\label{fig:triageguard}
\end{figure*}

Algorithm~\ref{alg:tg} links ranking quality to recovery time. A failure
on a pre-spared module switches to its hot spare in
$t_{\mathrm{sw}} = 5$ ms, while a failure anywhere else draws from the
shared pool at $t_{\mathrm{miss}} = 150$ ms. Call the share of expected failure probability that lands on protected
modules the \emph{intercepted mass} $\mu$.  The mean time to recover
(MTTR) is $\mu\, t_{\mathrm{sw}} + (1-\mu)\, t_{\mathrm{miss}}$, so MTTR
falls linearly as the ranking intercepts more of the mass. Uniform
sparing of $B$ modules out of $M$ intercepts $B/M$ of it on average, and
a risk ranking improves on that only as far as it aims at the modules
that will fail. The all-dynamic baseline, in which every failure draws
from the shared pool, bounds the MTTR of any ranking from above, because a protected failure
can only shorten a repair.

Fig.~\ref{fig:triageguard} runs this procedure on the shape of
Section~\ref{sect:odp}, 18 field-replaceable modules of 8 lasers each,
with hot spares for 4 of the 18; its left panel is a worked example at
$\tau = 0.35$. The
simulation draws each module's true risk from its temperature, age, and
FEC bin, fails each module with a probability that rises with that risk,
and pre-spares the four modules ranked highest. The ranking sees the
true risk through Gaussian error with a standard deviation of 0.30 on
stale telemetry and 0.05 on telemetry that ThermaLead refreshes each step. An oracle ranks on the
true risk itself. Over 30 seeds of 400 episodes, with every policy
scored on the same draws, the stale ranking intercepts 0.264 of the
failures, the fresh ranking 0.307, and the oracle 0.311. Uniform sparing
intercepts 0.222, close to its expected $4/18$. Averaged over the seeds, mean time to recover falls by 5.1\%,
10.4\%, and 10.9\% against uniform sparing, each within about one point
at 95\% confidence. Fresh telemetry thus recovers nearly all of the
oracle's margin, and that margin is modest at a budget of 4 modules in
18. The two error levels are assumptions, not measurements. Field
rankings suggest such concentration is attainable: on a production GPU
fleet, a learned risk ranking catches 64\% of future failures in its top
5\% of nodes \cite{nr_ma26}, and telemetry-based failure prediction for
optical transceivers reaches an F1 score of 0.89 on production data
\cite{nr_wang26b, nr_su26}. Neither measures the intercepted mass of a
co-packaged module pool. Under these assumptions the fresh telemetry
ThermaLead supplies doubles the cut a stale ranking achieves, and a
field measurement of the intercepted mass is what would confirm it.

\section{Performance Evaluation}
\label{sect:perf}

This section turns the ceiling of Section~\ref{sect:ceil} into design
guidance and then evaluates the integrated control plane.
Theorem~\ref{thm:hier} says where in the pool to decorrelate,
Observation~\ref{obs:split} prices separation against spare lasers, and
Observation~\ref{obs:rhoinv} shows that the prescription holds across
the plausible band of the correlation. Theorems and propositions are
proved, apart from the steps their proofs mark as numerical checks; an
observation is a computed finding under the cost model stated below.
Section~\ref{sect:corfaccomp} reports the integrated goodput, its
ablation, and its tail, and Proposition~\ref{prop:quantile} gives a
closed form for the part of that tail the correlation drives.  The
relevant software implementation of these components and the hardware
realization is available at \cite{implzip26}.

\subsection{Hierarchical Correlation Ceiling}
\label{sect:hier}

Lowering the correlation raises the ceiling, but a single pool-wide
correlation gives a designer nothing to act on, because it does not say
which lasers to pull apart from which. Section~\ref{sect:ceil} gave
every pair of lasers the same correlation $\rho$. Deployments are not
built that way. The lasers sit in field-replaceable modules, and a laser
shares more with the lasers in its own module than with the rest of the
pool. We model a shape of 18 modules of 8 lasers, the module count and
module size that NVIDIA's figures imply for one Quantum-X switch
(Section~\ref{sect:odp}). The active channels
and spares of Section~\ref{sect:lr} are drawn from a population of this
kind. A module is the unit of thermal and power isolation, and separate
modules are kept apart by design. Two lasers in the same module therefore fail together
with correlation $\rho_{\mathrm{within}}$, and two lasers in different
modules with the smaller $\rho_{\mathrm{across}} < \rho_{\mathrm{within}}$.
Reliability engineering treats common causes in the same grouped way.
The alpha-factor and multiple-Greek-letter models of nuclear safety
analysis apply a common cause within a common-cause component group and
treat separate groups as independent \cite{nureg5485}, which is the
limiting case $\rho_{\mathrm{across}} = 0$ of the model here.

\begin{figure}[!t]
\centering
\includegraphics[width=\linewidth]{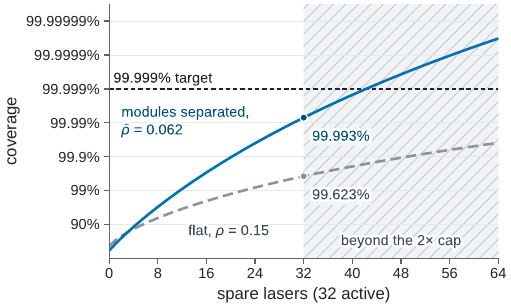}
\caption{Hierarchical ceiling for the 32-channel pool. The curves plot coverage against the number of spare lasers. The dashed curve is a flat pool in which every pair of lasers has correlation $\rho = 0.15$. The solid curve is the same pool with the average correlation of the modeled $18 \times 8$ module structure, $\bar{\rho} = 0.062$, evaluated with the flat model [derived].}
\label{fig:hier}
\end{figure}

Consider a pool of $M$ modules holding $L$ lasers each, in which a single laser fails with marginal probability $\lambda$, pairs in the same
module are correlated at $\rho_{\mathrm{within}}$, and pairs in
different modules at $\rho_{\mathrm{across}}$, with
$\rho_{\mathrm{within}} \ge \rho_{\mathrm{across}} > 0$. Concretely, a pool-wide level $\Theta_0 \sim \Beta(a,b)$ is drawn with
$a$ and $b$ set by $\lambda$ and $\rho_{\mathrm{across}}$; each module
$j$ then draws its own level $\Theta_j$, independently given $\Theta_0$,
from a beta law with mean $\Theta_0$ and intraclass correlation
$(\rho_{\mathrm{within}} -
\rho_{\mathrm{across}})/(1-\rho_{\mathrm{across}})$; and the lasers of
module $j$ fail independently at level $\Theta_j$.

\begin{theorem}[Hierarchical ceiling]

For the hierarchical system described above, the following statements
hold true:

\begin{enumerate}
  
\item As the module size $L$ grows with $M$ fixed, the failure fraction $F/N$ has limiting variance
  $\lambda(1-\lambda)[\rho_{\mathrm{across}} + (\rho_{\mathrm{within}} -
  \rho_{\mathrm{across}})/M]$, so the within-module term is suppressed
  by $1/M$, and as modules are added the across-module correlation sets
the floor.
  
\item As the module size $L$ grows with $M$ fixed, the failure
  fraction converges to the average $\bar{\Theta} =
  \frac{1}{M}\sum_{j=1}^{M}\Theta_j$ of the per-module failure levels,
  so availability converges to a ceiling $\Pr[\bar{\Theta} \le \sigma] < 1$, exactly as in
Theorem~\ref{thm:ceil}.

  \item The flat model of Section~\ref{sect:ceil} is the exact
    special case $\rho_{\mathrm{within}} = \rho_{\mathrm{across}}$.

    \end{enumerate}
    \label{thm:hier}
\end{theorem}

By Part~(1), adding modules averages away the correlation inside a
module and leaves the correlation between modules untouched, so a
designer gains more from separating modules than from separating the
lasers within one.

\begin{proof}[Proof sketch]
Part~(1) is the law of total variance applied to the two exchangeable
levels, and the factor $1/M$ appears because the within-module component
is averaged over $M$ module draws that are independent given $\Theta_0$. Part~(2) repeats the
law-of-large-numbers argument of Theorem~\ref{thm:ceil} one module at a
time: condition on the per-module levels $\Theta_1, \dots, \Theta_M$ (indexed
by $j$),
and the pool's failure fraction converges to their average. Part~(3) is
immediate, since equal correlations collapse the two levels into a
single beta-binomial, which reproduces the flat model of Section~\ref{sect:ceil} digit for digit.
\end{proof}

The pool-average pairwise correlation is
\[
\rhobar = \frac{n_{\mathrm{same}}\,\rho_{\mathrm{within}}
  + n_{\mathrm{cross}}\,\rho_{\mathrm{across}}}{n_{\mathrm{same}} + n_{\mathrm{cross}}},
\]
where $n_{\mathrm{same}}$ counts the pairs inside
one module and $n_{\mathrm{cross}}$ the pairs spanning two. There are
$n_{\mathrm{cross}} = \binom{M}{2}L^2$ cross-module pairs against $n_{\mathrm{same}} = M\binom{L}{2}$ same-module pairs, a
ratio of $(M-1)L/(L-1)$ that is close to the module count $M$. At the
$18 \times 8$ shape that is 9,792 cross-module pairs against 504
same-module ones, roughly 19 to 1, so $\rhobar$ tracks the across-module
correlation almost entirely. Evaluated at $\rhobar$, the flat,
single-level beta-binomial model of Section~\ref{sect:ceil} matches the
tail of the true nested process to within about one percentage point
from 8 spares up, and to within 0.63 points from 12 spares up. This shortcut, the flat model evaluated at $\rhobar$, is an
approximation, so wherever we use it we also report the nested value.

One objection to co-packaging is that it \emph{increases} correlation,
and it does, but only within a module. By Theorem~\ref{thm:hier} the
ceiling rises with separation \emph{across} modules, and the field-replaceable faceplate modules of the architecture provide that
separation. Set $\rho_{\mathrm{within}} = 0.30$ and $\rho_{\mathrm{across}} = 0.05$,
and the pool-average correlation of the $18 \times 8$ shape becomes
$\rhobar = 0.062$. Applied to the 32-channel pool at the $2\times$
budget cap, with the spare count unchanged, the shortcut lifts coverage
from the 99.623\% of the flat model at $\rho = 0.15$ to 99.993\%
(Fig.~\ref{fig:hier}), well past the 99.670\% that even a large flat
pool approaches. The nested model confirms the gain: intrepreted as 8 modules
of 8, the 32-channel pool has its own $\rhobar = 0.078$, at which the
shortcut gives 99.975\% and the nested tail 99.985\%, so the shortcut
errs low. The full $18 \times 8$ shape at its own $2\times$ cap reaches
99.997\%.

\subsection{Spares Versus Separation}
\label{sect:split}

How far to take separation depends on prices, and once prices are attached the choice between
decorrelating and over-provisioning becomes an optimization with a
definite answer. We state the prices as assumptions: \$1,500 per spare laser, since we found no public price for pluggable
laser sources, and a one-time separation
premium of 20\% of the active-laser capital to move a pool from poor
separation to good separation. Neither is a quote. The premium stands in
for second sourcing and for independent power and cooling, which IEC~61508-6 rewards with a lower beta factor \cite{iec61508-6} but does not
price. Every dollar figure below scales with the per-laser price, so the
ratio between the two paths does not depend on it.

\begin{observation}[Optimal split]
Under the stated cost model, given a budget spendable on either spares
(raising the spare ratio) or separation (lowering $\rho$ at the stated
premium), the minimum-cost path to a 99.999\% coverage target at the
stress operating point decorrelates the pool first and buys spares second.
\label{obs:split}
\end{observation}

\begin{figure}[!t]
\centering
\includegraphics[width=\linewidth]{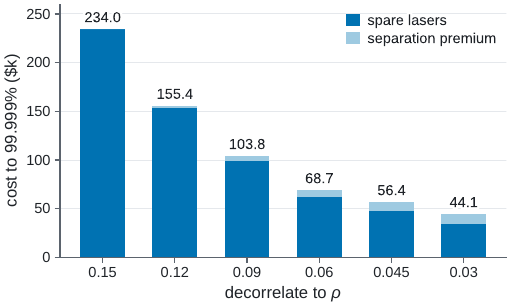}
\caption{Optimal split. Under the stated cost model, decorrelating first and buying spares second reaches 99.999\% coverage at a $5.3\times$ saving over spares alone. Each bar stacks the two costs of reaching the target, the spare lasers bought and the engineering premium paid for physical separation, and the number above it is the total, in thousands of dollars [derived; stated cost model].}
\label{fig:split}
\end{figure}

Fig.~\ref{fig:split} prices two paths to the same target. A candidate
plan is a pair: a spare count $s$, and a decorrelation target
$\rho'$ at or below the pool's starting correlation $\rho_0$. Its cost
is $s$ times the per-laser price plus the separation premium for
reaching $\rho'$. Its
availability comes from the same beta-binomial coverage model that
Algorithm~\ref{alg:lr} uses, and the reported point is the cheapest plan
that meets the 99.999\% target. Both paths are priced on coverage
alone (spares absorbing every failure), with no credit for the fast
recovery of Section~\ref{sect:tg}. Reaching 99.999\% coverage by spares
alone at $\rho = 0.15$ takes 156 spare lasers, about \$234,000.
Decorrelating to $\rho = 0.03$ instead costs about \$9,600 in separation
premium, which is the 20\% figure applied to the \$48,000 of
active-laser capital in the 32-channel pool. That leaves only 23 spare
lasers to buy, at \$34,500, for \$44,100 in total. Under the stated cost model, separating first and sizing second therefore reaches the
same target at a $5.3\times$ saving over spares alone. At this premium
the ordering is not close: it would reverse only if separation cost
several times the pool's entire laser capital. The split depends on the
separation-premium assumption, and a reader with a different premium can
re-solve it with the same coverage model and prices.

\subsection{Optimality with Varying \texorpdfstring{$\rho$}{rho}}
\label{sect:rhoinv}

The plan priced above starts from an assumed correlation, and $\rho$ has
not been measured, so the analysis is weakest here. We do not defend a particular value of $\rho$.

\begin{observation}[$\rho$-invariance]
Under the stated cost model, for every starting correlation in the
plausible band $\rho_0 \in [0.08, 0.35]$ the minimum-cost path to the 99.999\% coverage target decorrelates the pool first, toward $\rho \approx
0.03$. The optimal action never reverses direction within the band.
\label{obs:rhoinv}
\end{observation}

\begin{figure}[!t]
\centering
\includegraphics[width=\linewidth]{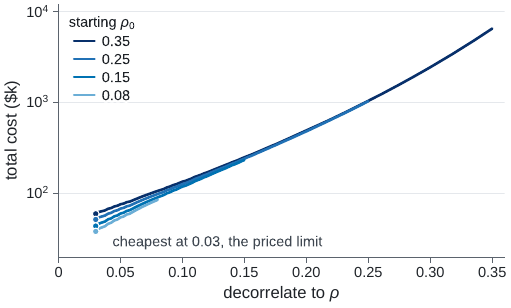}
\caption{$\rho$-invariance. The curves plot the total cost of reaching the coverage target against how far the pool is decorrelated, one curve for each starting correlation $\rho_0$. Every curve falls all the way to $\rho = 0.03$, the best separation the premium is priced for, so the cheapest plan decorrelates as far as the model allows, wherever the pool starts within the plausible band [derived].}
\label{fig:invariance}
\end{figure}

Fig.~\ref{fig:invariance} sweeps the starting correlation across the
band, with a partial move priced in proportion to the correlation it
removes: \$800 for each 0.01, so the move from 0.15 to 0.03 costs the
\$9,600 above. A measured value of $\rho$ will therefore set the size of the saving,
not the choice of action.

\subsection{Performance of the Integrated System}
\label{sect:corfaccomp}

Each component has been evaluated alone against the baseline it
replaces, and here their results are composed into the effective cluster
goodput of Section~\ref{sect:intro}.

\begin{figure}[!t]
\centering
\includegraphics[width=\linewidth]{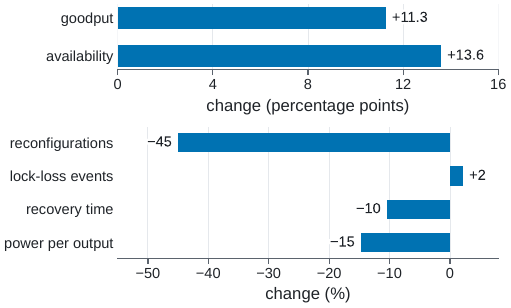}
\caption{The full control plane against the deployed baseline: one bar per system-level KPI at the stress operating point, each showing the change from the baseline. Under the coverage reading, goodput rises 11.3 points and pool
availability 13.6 points with a 49-spare pool past the $2\times$ cap
(11.1 and 13.3 points held to the cap). Reconfigurations, recovery time,
and the power needed per unit of training output fall. Lock-loss events rise 2\% [composed].}
\label{fig:kpi}
\end{figure}

At the stress setting of Section~\ref{sect:analysis}, composing the
simulated component results lifts effective cluster goodput from 65.2\%
to 76.5\%, a rise of 11.3 points, or 17.3\% in relative terms
(Fig.~\ref{fig:kpi}). The full-stack figure is composed by arithmetic on
the component results and is not a separate emulation: the 65.2\%
baseline, a calibration anchor rather than a simulated value, is
propagated through the simulated availability, lock-loss, and
reconfiguration changes under
\[
\text{goodput} = \text{ideal} \times \text{availability}
  \times (1-\text{thermal}) \times (1-\text{reconfig}).
\]
Here $\text{availability}$ is the laser-pool
availability of Section~\ref{sect:lr}, $\text{thermal}$ is the share of goodput lost to lock-loss stalls, and
$\text{reconfig}$ is the share lost to switch reconfigurations. The laser pool enters through its per-window coverage, which depends on
the window only through $\lambda$.

That availability factor carries an assumption that sets most of the
result. The coverage reading counts a window whose failures exhaust the
spares as lost from start to finish. If only the dark channels'
bandwidth is lost instead, the availability factor becomes the expected
lit share of the 32 channels; we call this the lit-share reading. That
share is 98.0\% for the four-laser reserve, against its 86.3\% coverage,
and 99.98\% for the full stack. With the 65.2\% anchor held fixed, that
reading lifts the full stack only to 67.4\%, a rise of 2.2 points, or
3.4\% relative. Each reading implies its own ideal goodput. A real job
need not fall between the two. When in the window the spares run out and for how long also move the
figure, and this work models neither. Within the window model, how a
job handles a dark channel is one parameter: if, once the spares run
out, it cannot use a share $c$ of its remaining bandwidth, the
availability factor is $c$ times the coverage plus $1-c$ times the lit
share, and the gain rises from 2.2 points at $c = 0$, a fabric that routes
around a dark channel, to 11.3 at $c = 1$, a synchronous job halted until
the restock. Every figure below uses $c = 1$, the coverage reading.

Each component reports its result in its own units, events or
milliseconds, and the baseline's two loss shares, $\text{thermal}$ and
$\text{reconfig}$, convert the results into goodput. The full stack keeps each share in
proportion to its simulated count, 472/462 for lock-loss events against
the dither lock and 8.8/16 for the proxy count of reconfigurations, and recovery time has no
term in the composition. We have no measured costs, so we set both
baseline shares at three percent, and Section~\ref{sect:measure} names
them among the cheaper things to ground on real hardware. Sweeping the
two shares by a factor of two in each direction moves the baseline
across 61.2--67.2\% and the full stack across 72.8--78.3\%, while the
full stack's gain relative to the baseline stays within 16.5--18.9\%.
The relative gain is therefore robust to the two loss shares, though not
to the reading of a shortfall, and the absolute pair should be read with
its band. The released composite model computes coverage and lit share from the
model of Section~\ref{sect:ceil} and produces every number in this
section.

\begin{figure}[!t]
\centering
\includegraphics[width=\linewidth]{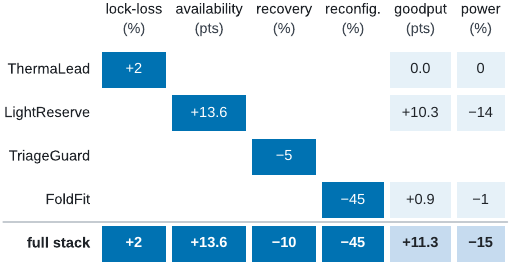}
\caption{Ablation. The first four rows each run one component alone and
the last row runs the full stack; each entry is its change from the
deployed baseline. Among the four layer KPIs, a component alone moves
only the one its layer owns: lock-loss for ThermaLead, availability for
LightReserve, recovery time for TriageGuard, and reconfigurations for
FoldFit.
TriageGuard alone runs on stale telemetry, since ThermaLead supplies the
fresh signal, and its recovery-time cut has no term in the goodput
formula, so its goodput and power entries are blank. The availability entry uses a 49-spare pool past the $2\times$ cap,
under the coverage reading. LightReserve and FoldFit each add goodput
and save power per unit of output. ThermaLead moves goodput by less than
0.1 point and power by less than 0.1\% [composed].}
\label{fig:ablation}
\end{figure}

Fig.~\ref{fig:ablation} isolates each component's contribution to the
key performance indicators (KPIs). LightReserve contributes the most,
because availability dominates goodput once failures correlate.
Composing the availability gain alone, from 86.3\% to 99.9\%, moves
goodput from 65.2\% to 75.5\%. Credited first, availability therefore accounts for about 10.3 of the
11.3 points, and the reconfiguration factor adds the remaining 1.0,
since the thermal factor barely moves. Credited the other way round, the
two factors add 0.9. The availability share is
provisioning: it comes from the spare lasers that the coverage bound
sizes, 49 of them for 99.9\% at the stress point, past the $2\times$ cap
of 32. Held to the cap, the pool reaches 99.623\% and composed goodput 76.3\%,
against 76.5\% uncapped.

\subsubsection{A Correlated-Factor Composition}

The multiplicative model treats availability and thermal loss as
independent, but a shared cause could move both. The coupled model
assumes that one thermal event drives both the lock-loss stalls
(ThermaLead's domain) and the correlation $\rho$ (LightReserve's), so it
moves both factors through the latent variable $\Theta$ of
Theorem~\ref{thm:ceil}.
Multiplying the two means gives an expected value and understates the
tail, where both factors fail together. The coupled model draws $\Theta
\sim \Beta(a,b)$ for each window. Given the draw, the failures among the
full stack's lasers are binomial at that level, so the pool either covers every one of them or runs out. A window that
runs out counts as lost, as the coverage reading assumes. The
thermal loss share is the full stack's share scaled by $\Theta/\lambda$.
Coverage and thermal loss thus keep their stated means, and the model
reports the distribution of the goodput each window realizes.

\begin{figure}[!t]
\centering
\includegraphics[width=\linewidth]{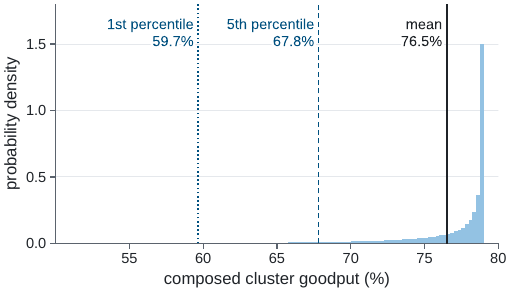}
\caption{Correlated-factor composition. Histogram of composed goodput, under the coverage reading, when the
common driver $\Theta$ moves availability and thermal loss together.
Vertical lines mark the mean (76.5\%), the fifth percentile (67.8\%),
and the first percentile (59.7\%). The mean is within 0.1 point of the
independent model's point estimate, which has no tail. Below
the plotted range fall 0.1\% of windows, most of them windows whose
failures exhaust the spares [composed].}
\label{fig:coupling}
\end{figure}

The means agree: 76.5\% under the coupled model and for the independent
point estimate. The coupling changes the tail (Fig.~\ref{fig:coupling}).
Goodput at the fifth percentile is 67.8\%, and at the first percentile
it is 59.7\%. A cluster
provisioned to the mean therefore misses its plan by 8.7 points or more
in one window out of twenty, and by 16.8 points or more in one out of a
hundred. This spread comes from the $\rho$ of Theorem~\ref{thm:ceil},
and the decorrelation that raises the ceiling also tightens the spread.
None of the per-component guarantees composes into a worst-case
bound on effective cluster goodput, but part of the tail can be proved
(Section~\ref{sect:tail}).

\begin{figure}[!t]
\centering
\includegraphics[width=\linewidth]{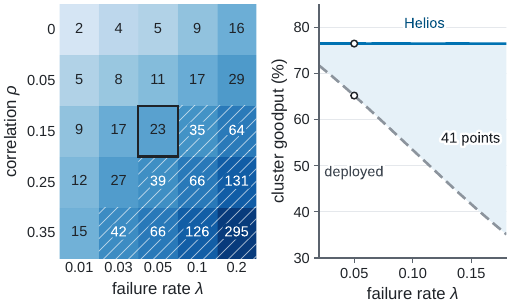}
\caption{The swept range. Left: the spares needed for 99\% coverage of 32 active channels as the failure rate $\lambda$ and the correlation $\rho$ vary. Hatched cells need more than the $2\times$ cap of 32 spares, and the outlined cell is the stress operating point. Right: goodput of the deployed baseline (dashed) and of the full stack,
its pool sized to 99.9\% coverage at each rate, as the failure rate
grows at $\rho = 0.15$, with circles at the stress operating point. That sizing already passes
the $2\times$ cap at the stress point. Shading marks the gap between the
two curves. Both figures use the coverage reading [composed].}
\label{fig:range}
\end{figure}

The stack is worth more under worse conditions. At 50~FIT the four-laser reserve covers 99.974\% of 90-day windows and
99.896\% of one-year windows. The 90-day figure already exceeds the 99.9\% target, so availability is
held at that value. With the thermal and reconfiguration factors at
their simulated values, the composed gain falls to about 1.0 point, or
1.3\% relative, all of it from the reconfiguration factor.
As the failure rate rises at $\rho = 0.15$, the baseline's goodput drops
sharply while the stack, its pool sized to 99.9\% coverage, holds, and
the gap between them widens to 41 points at a failure rate of 0.18 per
window (Fig.~\ref{fig:range}). At that rate the target takes 116 spares,
far past the $2\times$ cap; held to the cap, the stack keeps 73.2\% against the baseline's 35.1\%, a
gap of 38 points.

\subsection{What Can Be Proved About the Goodput Tail}
\label{sect:tail}

Section~\ref{sect:corfaccomp} reported a fifth-percentile goodput of 67.8\%, obtained by sampling the
common driver $\Theta$. A sampled
percentile is a simulation output: run the simulation differently and
the number moves. The systematic part of goodput, the part $\Theta$
moves, has quantiles in closed form, so that part of the tail can be
reasoned about as well as observed. The mechanism is the standard monotone transport of quantiles. 

\begin{figure}[!t]
\centering
\includegraphics[width=\linewidth]{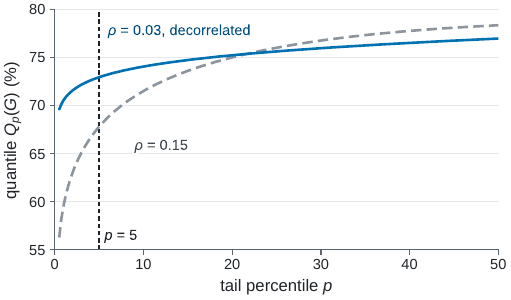}
\caption{Quantile transport. Closed-form quantile of the
correlation-driven part of goodput, $G(\Theta)$ of the coupled model in
Section~\ref{sect:corfaccomp}, against the tail percentile $p$, in
percent. The dashed curve is the $\rho = 0.15$ operating point and the solid curve the same pool after decorrelation to $\rho = 0.03$. Decorrelation lifts the lower tail, including the fifth percentile; the curves cross near $p = 22$, above which the more correlated pool sits higher [derived].}
\label{fig:quantile}
\end{figure}

\begin{proposition}[Quantile transport]
Let $G$ be the systematic part of the composed goodput, a strictly
decreasing function of the single driver $\Theta \sim \Beta(a,b)$; the
realized goodput is $G$ plus the conditional binomial variation of
failures given $\Theta$. Then, for any $p$, the $p$-quantile of $G$ is
$Q_p(G) = G\bigl(F_\Theta^{-1}(1-p)\bigr)$. For $p$ below a threshold $p^{*}$ that depends on $\lambda$ and $\rho$
and that we locate numerically, $\partial Q_p/\partial \rho < 0$, so
decorrelation lifts that part of the tail.
\label{prop:quantile}
\end{proposition}

\begin{proof}[Proof sketch]
The identity is monotone transport of quantiles: a strictly decreasing
map sends the $(1-p)$-quantile of its input to the $p$-quantile of its
output. The sign follows because, for $p < p^{*}$, raising $\rho$ moves the
$(1-p)$-quantile of $\Theta$ up, which lowers $G$ there; $p^{*}$ is
where that quantile stops rising with $\rho$, which we locate
numerically. We
checked the identity numerically, comparing empirical quantiles of four million Monte Carlo draws of
$G(\Theta)$ against the closed form, and the two agree to within $2
\times 10^{-4}$ at the reported percentiles.
\end{proof}

Located numerically at $\lambda = 0.05$, $p^{*}$ falls from the 31st
percentile at $\rho = 0.03$ to the 15th at $\rho = 0.15$ and the 8th at
$\rho = 0.35$, so it covers the fifth percentile across the plausible
band.

Fig.~\ref{fig:quantile} plots this closed form. Realized goodput varies
further: the conditional binomial spread of failures given $\Theta$
widens its distribution in both directions. Here $G$ is the goodput expected at a given $\Theta$, a mean and not a
minimum. In the coupled model of Section~\ref{sect:corfaccomp}, a window's realized goodput sits slightly above $G$ when its spares
cover every failure. It falls to zero in the 0.1\% of windows where they
run out, so $Q_p(G)$ is no floor under realized goodput. At the percentiles reported there the two agree to within 0.02 points:
at $p = 5$ the closed form gives the 67.8\% sampled there. Turning $Q_p(G)$
into a floor under realized goodput would require bounding the
conditional binomial spread as well, which we do not attempt here.

\subsection{Simulation Methodology and Reproducibility}
\label{sect:simmeth}

The simulated and composed results in this and the preceding sections
come from a calibrated simulation stack written in Python with NumPy and
SciPy and run with fixed seeds; the derived results are closed-form
evaluations. None of them is a hardware measurement. Each figure caption
carries a provenance tag: [derived] for closed-form evaluation, [composed] for results composed
arithmetically from the component outputs, such as goodput and recovery
time, [schematic] for architecture drawings, and
[synth; pnr; verified] for the synthesis, place-and-route, and
verification outputs reproduced in Section~\ref{sect:hwimp}. Some
captions combine or qualify these base tags, as in [derived; stated cost model], [synth; pnr], and [structure matches the synthesized
sources].

The configuration of each component follows, for reproduction;
Section~\ref{sect:cpfrwk} reports the results.
LightReserve runs Algorithm~\ref{alg:lr}. The 32-channel calibration
uses independent failures at $\lambda = 0.05$, $0.10$, and $0.20$ with a
99\% coverage target, and a correlated case at $\lambda = 0.05$ and $\rho = 0.15$ with targets of 99\%, 99.9\%, and 99.999\%. FoldFit is simulated on block-structured MoE demand, dense within each
expert block of 8 engines and sparse across blocks, on 8 channels at 8
to 128 engines. The simulation binary-searches the deadline and, at each
candidate, places the jobs longest first on a channel that already holds
the job's expert block when one has room, and otherwise on the
least-loaded channel with room. It omits the port condition, and with it
the alignment step, so makespans are measured against the larger of the largest demand and
the average load, which they meet from 64 engines up. ThermaLead runs on the heat-cycle trace of Section~\ref{sect:tl}: over
30 seeds the simulation counts lock-loss events per 2000 control steps
for ThermaLead, its gated variant, the dither lock, and an oracle;
it also sweeps the share of steps that carry a noise burst from 30\%
down to none. TriageGuard is configured as in Section~\ref{sect:tg}. 

A reproducibility bundle accompanies the paper. It holds the
register-transfer sources, testbenches, proofs, build scripts, and
130~nm layout flow behind Section~\ref{sect:hwimp}; the component simulator behind the results of Section~\ref{sect:cpfrwk},
with the ThermaLead baseline and gate studies and the abrupt-transient
sweep of Section~\ref{sect:limits}; the composite model, which reads the
simulator's output, behind the goodput, recovery-time, and capital
figures; the cross-workload simulations of
Section~\ref{sect:transfer}; and the scripts that evaluate the
closed-form results and draw the figures. 
\section{Hardware Implementation of the Control Plane}
\label{sect:hwimp}

Halley's fallback to reactive control, which overrules a forecast that
is stale or unsure, needs a circuit to enforce it. We implemented the
two safety-critical parts of \textsf{Halley} as one design for a Lattice
iCE40UP5K \cite{ref47}, a small field-programmable gate array (FPGA).
The first is the fail-safe watchdog, which enforces the fallback rule of
Section~\ref{sect:cpfrwk}: cycle by cycle, it decides whether the
predictive or the reactive command reaches the actuator. The second is
ThermaLead's local thermal servo, which drives the per-ring heaters and
is the only one of the four controllers that closes a loop at hardware
speed. LightReserve, FoldFit, and TriageGuard decide on slower time
scales: LightReserve sizes a pool when it is provisioned or restocked,
FoldFit computes a schedule for each demand matrix, and TriageGuard
ranks a pool's modules from their telemetry. A supervisory processor,
such as the supervisory host of Section~\ref{sect:pkg}, can run all
three in software, so none of them needs dedicated logic. All four need
telemetry and actuation that published laser and switch interfaces
provide only in part (Section~\ref{sect:losses}), which is why the
fallback is the part of \textsf{Halley} that can be built and measured today.

The design was synthesized, placed and routed, and formally verified
with an open toolchain pinned to fixed versions: Icarus Verilog~12.0
\cite{ref44}, Yosys~0.33 \cite{ref45}, nextpnr~0.6 \cite{ref46}, and the
icetime timing analyzer of Project IceStorm. The method is test-driven
\cite{ref48}, with an executable specification written before the
circuit. Every resource figure below regenerates exactly from the
released register-transfer sources. The timing figures are the recorded
results for each placement seed; a rebuild can land at different
frequencies on a given seed, and in our independent rebuild of the core,
the servo, the eight-instance layout, and the ECP5 port, every seed we
rebuilt still closed timing.

\subsection{Circuit Design}

\begin{figure*}[!t]
\centering
\includegraphics[width=\linewidth]{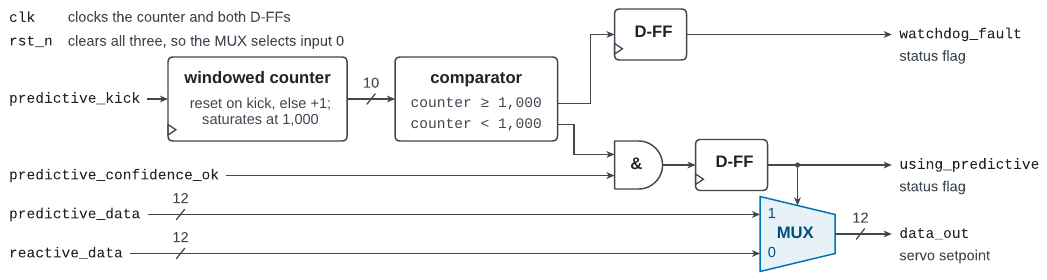}
\caption{The fail-safe watchdog and bypass multiplexer (MUX); \&{}
marks an AND gate and D-FF a register. The counter stops at the
timeout, a test that synthesis shares with the comparator [structure
matches the synthesized sources].}
\label{fig:watchdog}
\end{figure*}

\begin{figure*}[!t]
\centering
\includegraphics[width=\linewidth]{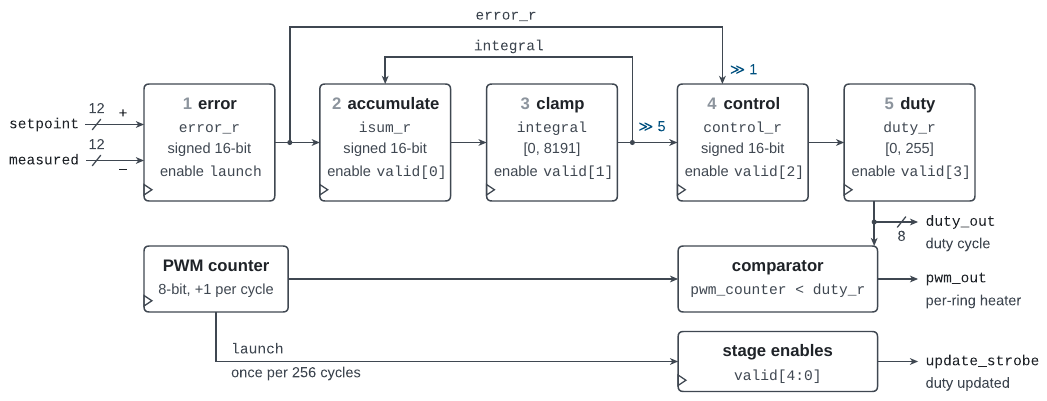}
\caption{The ThermaLead local servo. The five numbered blocks are the
pipeline stages of a multiplier-free proportional-integral law, from
temperature error to heater duty cycle. A label $\gg n$ means an
arithmetic shift right by $n$ bits, which applies a gain without a
multiplier. The strip below is the 256-cycle pulse-width-modulation
generator, which drives the per-ring heater and, once per period,
launches the stage enables that step one update through the pipeline
[structure matches the synthesized sources].}
\label{fig:servo}
\end{figure*}

The safety core is one circuit with two halves that share a clock and a
reset. The first half is the watchdog of Fig.~\ref{fig:watchdog}, module
\texttt{watchdog\_bypass}. It lets the predictive path drive the
actuator only while the predictor keeps answering and reports
confidence. A windowed counter restarts on each
\texttt{predictive\_kick} from the predictor and otherwise counts up
until it saturates at the 1,000-cycle timeout, which a magnitude
comparator detects. A registered \texttt{using\_predictive} flag is set
when the counter is in window and \texttt{predictive\_confidence\_ok} is
high, and a second register raises \texttt{watchdog\_fault} one cycle
after the window closes. The flag is the select of an output multiplexer
that holds no state of its own, and reset and power-up clear it, so the
reactive path is chosen unless the flag is set. A crash, hang,
low-confidence signal, or reset therefore parks the actuator on the
reactive path, which is always live, within at most one trust window,
20~\textmu s at the 50~MHz target clock, plus one clock after the last
kick, 20.02~\textmu s in all. After a fresh kick, the predictive path is
restored within two clock cycles.

The second half is ThermaLead's local loop, module \texttt{thermal\_pi}
(Fig.~\ref{fig:servo}), which turns a measured temperature error into
heater drive. Five pipeline stages compute error, accumulate, clamp,
control, and duty. Together they implement a proportional-integral law
whose two gains are arithmetic shifts, one bit for the proportional gain
and five for the integral gain. The law therefore needs no multiplier,
and it maps onto look-up tables (LUTs) and carry cells instead of a
dedicated digital-signal-processing multiplier. The accumulator, clamped
to $[0, 8191]$, is the only state carried from one update to the next.
The control sum is clamped to $[0,255]$ and registered as the duty. A
separate free-running 8-bit generator holds \texttt{pwm\_out} high while
its counter sits below that duty, and that signal drives the per-ring
heater. The law updates once per 256-cycle pulse-width period,
5.12~\textmu s at 50~MHz, so the five-cycle pipeline latency is
negligible against that period, and the duty stream is bit-identical to
a single-cycle implementation of the same law. At that rate the heater
drive does not limit ThermaLead's control step; the heater's thermal
response does (see the closed-loop sweep below).

\subsection{Chip Design}

Table~\ref{tab:fpga} gives the toolchain results. The watchdog synthesizes to 36 LUTs and the servo to 97, and
the combined core is exactly their sum, 133 LUTs, so either half can be
budgeted on its own. Counted at synthesis for the core alone, the 133
LUTs are 2.52\% of the device's 5,280.

Under nextpnr, timing closes on every placement seed. Placed and routed
behind the on-chip serial test harness, the combined core closes at
68.97--74.69~MHz across ten seeds, a worst-case margin of 38\% over the
target. The independent icetime analyzer
puts the combined core at 71.33--74.79~MHz on the same placements, so
both models clear the target on every seed. 

The core is paired with a production-grade sensor path on a mode-zero
serial-peripheral-interface (SPI) link. With that path the design comes to 161 LUTs and
closes on 10 of 10 seeds under both models. The same sources also port
to a second FPGA, a Lattice ECP5-25, where they close at 196.97--211.91~MHz across three seeds, so the result
is not tied to one device. Unchanged, the core and its serial test
harness also lay out with LibreLane~3.0.14 as a 142 by 153~\textmu m
SkyWater 130~nm block, 0.022~mm$^2$, with 128 flip-flops, the core's 90
and the harness's 38, among 995 standard cells besides fill
(Fig.~\ref{fig:device}b). The block meets 50~MHz at all nine signoff
corners, with 11.0~ns of worst-case setup slack, and passes design-rule,
layout-versus-schematic, and antenna checks; it has not been fabricated.

\begin{table}[!t]
\centering
\footnotesize
\setlength{\tabcolsep}{4pt}
\begin{tabular}{@{}lrrrl@{}}
\toprule
Module & LUTs & FFs & Carries & $f_{\max}$ (MHz) \\
\midrule
\texttt{watchdog\_bypass} & 36 & 12 & 14 & 92.54 \\
\texttt{thermal\_pi} & 97 & 78 & 53 & 81.98--87.01 \\
Combined safety core & 133 & 90 & 67 & 68.97--74.69 \\
Core $+$ SPI sensor path & 161 & 149 & 68 & $>50$ \\
\bottomrule
\end{tabular}
\vspace{1.6ex}
\caption{Synthesized footprint and closed timing on the iCE40UP5K. Logic
counts come from Yosys \texttt{synth\_ice40}, and placement and routing
from nextpnr across 10 seeds (the watchdog alone across 3) against a 50~MHz target. FFs are
flip-flops, and $f_{\max}$ is the highest clock frequency at which timing closes.}
\label{tab:fpga}
\end{table}

Beyond the tests, four machine proofs by temporal induction in Yosys
check the core, each for every input. Temporal
induction proves a property outright for a bounded run of cycles, then
proves that any two consecutive cycles satisfying it force the next
cycle to satisfy it too, so the property holds for every cycle. The
first proof checks the watchdog against a one-bit-wider golden model, an
independently written reference circuit. The second proves the servo's
invariants: the accumulator stays in $[0, 8191]$, the integral and the
duty change only on their stage enables, and \texttt{pwm\_out} equals
the comparison of counter and duty. The third checks the servo pipeline
against the one-shot control law. The fourth checks the whole core,
watchdog and servo together, against an independent specification core
whose clamps are written as comparisons. The first two start checking
after eight cycles of warm-up, and the last two check every cycle from
power-on. All four report success on the induction step in the released
logs, so the bypass path that the control plane relies on is a proven
property of the register-transfer design.

The proofs show that the circuit computes the law it was meant to
compute. They do not show that the law is well behaved once the loop is
closed, so the servo was also run in closed loop against a second-order
thermal plant with quantization, noise, and one period of sensor delay.
Across a gain-and-pole sweep, it converges at all 31 feasible corners,
settling in 0.2 to 3.7~ms, and re-settles after a mid-run thermal step
in all 27 corners where one was injected. The one infeasible corner asks
for 500 counts at 0.6 times the nominal gain, more than the 480 the
plant reaches at full duty, and there the servo correctly holds the duty
at 255 and the integral at its cap. Its phase and gain margins stay at or above
22.5\textdegree{} and 14.9~dB across the sweep, and an independent
recomputation confirms them. The swept plant time constants include the 30.9~\textmu s and 0.71~ms of a measured microring heater \cite{ref49}
and the 13.8~\textmu s of a metal microheater \cite{ref50};
doped-silicon microheaters, at 0.98~\textmu s \cite{ref50}, are faster
than the sweep covers.

\subsection{Packaging}
\label{sect:pkg}

The fallback can be built from catalog silicon with an ordinary pinout:
the iCE40UP5K in the 48-pad SG48 package.

\begin{figure}[!t]
\centering
\includegraphics[width=\linewidth]{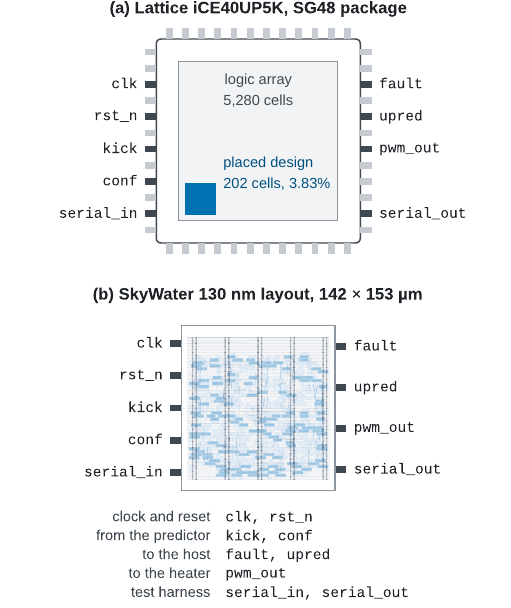}
\caption{The device, the placement, and the pins, in the iCE40 and 130~nm
builds. (a) Inside the 48-pad SG48 package of the Lattice iCE40UP5K, the inner square is the logic array, and the filled square, drawn to scale by area, is the placed design: 202 logic cells including the serial test harness. Pad positions and the design's place in the array are schematic
[synth; pnr]. (b) The signed-off SkyWater 130~nm layout, drawn from its
GDS: cells light blue, the 128 flip-flops darker, pins as laid out
[synth; pnr; verified].}
\label{fig:device}
\end{figure}

The interface brought out to the package is deliberately narrow
(Fig.~\ref{fig:device}). A deployed part would carry setpoints and
telemetry to the supervisory host one level above this device over a
management bus such as I2C, in place of the two-pin serial test port,
and the wide data buses stay on chip behind that serial test harness.
Placed together with the harness, the design occupies 202 of the 5,280 logic cells, or 3.83\%, against the
2.52\% that synthesis counts for the core alone. To test whether one device can serve several thermal domains, we placed
and routed eight instances of the core without the shared telemetry bus
that would report them to the supervisory host. They occupy 1,417 of the
5,280 logic cells (26.8\%) and close timing on all six seeds under both
models, at 55.25--58.59~MHz under nextpnr and 55.91--58.92~MHz under
icetime; the bus itself has not been placed.

Both safety-critical circuits are placed and routed for a shipping part,
close timing under the place-and-route tool on every seed, and use very
little area, and the bypass is a proven property of their
register-transfer design, so the fallback path can be built. Whether
Halley then delivers the reliability outcome still depends on the
unmeasured correlation $\rho$ of Section~\ref{sect:ceil}, which only the
field measurement of Section~\ref{sect:measure} would settle.
\section{Discussion}
\label{sect:disc}

\subsection{Impact on Capital and Power Costs}

The goodput gain of Section~\ref{sect:corfaccomp} is an upper bound on
the useful compute, and so the capital, that Halley frees. The coverage reading
multiplies the useful compute the same hardware delivers by up to 1.17,
a 17.3\% rise, and the lit-share reading by up to 1.03. Both factors
hold \emph{only if} fabric stall is the sole recoverable loss, and it is
not. Model FLOPs utilization (MFU), the share of peak arithmetic that a
training run keeps busy, also counts compute, memory, pipeline-bubble,
and data-loading losses that the control plane never touches. The
65.2\% goodput here and the 38--43\% MFU that Llama~3 reported
therefore do not conflict.

At that bound, the gain cuts the accelerators, capital, and
provisioned power needed for a given training output by up to 14.7\%,
or 3.3\% on the lit-share reading, for a fixed rack design and
power-smoothing scheme. Energy falls by less, because accelerators
waiting at a barrier draw less than full power \cite{ref24}. Where power
and local approval limit new AI data centers (Section~\ref{sect:intro}),
the gain becomes more training from existing sites instead of a saving. It is a
one-time factor on top of the 30 to 40\% yearly improvement in
accelerator performance per watt \cite{iea26keyquestions}, not a
substitute for it. At 50~FIT the same arithmetic starts from the 1.3\%
composed gain instead, and these capital and power figures fall with it.

A control plane can address only the fabric's share of recoverable
stall, the attribution fraction. Scaled by it, the coverage reading's
17.3\% rise becomes 13.0\%, 8.6\%, and 5.8\% at three-quarters, half,
and a third, and the lit-share reading's rise becomes 2.5\%, 1.7\%, and
1.1\%. Llama~3 gives the fraction a scale. A tenth of its 419 unexpected
interruptions were traced to network
switches, cables, and interface cards \cite{ref24}, which would put the
gain near 1.7\% on the coverage reading and 0.3\% on the lit-share
reading. That breakdown counts interruptions, not stalled time, so it
gives a scale rather than a value, and this work does not measure the
fraction for a co-packaged fabric.

\subsection{Addressing Practical System Losses}
\label{sect:losses}

The results bear on four problems that operators pay for today and that
deployed systems leave open.

The first is time lost at the synchronization barrier in hyperscale
pretraining. Llama~3 held effective training time above 90\% through its
419 interruptions \cite{ref24}. Laser-pool exhaustion would
add to that loss on a co-packaged fabric, and within a budget cap
Theorem~\ref{thm:ceil} shows that more spares alone cannot remove it.
Decorrelation with fast recovery is the alternative, and each point of
availability recovers training time in proportion to the
attribution fraction.

The second is the tail-latency objective of MoE inference. Generating a
token runs the same all-to-all exchange as training; that exchange takes
30--40\% of iteration time in DeepSeek-V3 training on H100 before
overlap with computation and under 5\% after \cite{ref37}. If decoding falls in the
same range, FoldFit's $11/9$ bound caps the worst-case inflation of an
iteration, under its stated conditions, at 6.7--8.9\% before switching delay and with circuits held across
steps, or at about 1\% once the exchange is overlapped.

The third is capacity planning before $\rho$ is measured.
Observation~\ref{obs:rhoinv} shows that decorrelating first is the
cheaper path across the plausible band, and the coupled model plans
against the goodput tail that the independent model hides.

The fourth is the control interface. The algorithms are cheap and step
aside on a flagged forecast, so apart from ThermaLead's forecast
(Section~\ref{sect:limits}) the obstacle to deployment is an interface
that exists only in part. External laser modules report per-lane bias,
power, and alarms and accept per-lane enable \cite{elsfpcmis}, and
commercial optical circuit switches expose NETCONF and RESTCONF control
\cite{polatis}. Neither the OIF management framework for co-packaged
engines nor the ELSFP interface, however, defines per-ring temperature
or lock state \cite{oifmgt, elsfpcmis}. One change would unblock both
deployment and the field measurement of $\rho$
(Section~\ref{sect:measure}): a published interface exposing per-laser
health and failure timestamps, per-ring temperature and lock state, and
switch reconfiguration hooks on the management interface the packages
already carry.

\subsection{Impact on Inference, Post-Training, and RL Traffic}
\label{sect:transfer}

Halley was developed for training, where the failure data of
Section~\ref{sect:bg} show the need. The same fabric now carries
inference and RL as well, but of the four components only FoldFit
carries a guarantee that transfers cleanly.

\textbf{FoldFit transfers:} MoE decoding runs the same all-to-all expert
dispatch on its critical path when experts are spread across the fabric. Serving that disaggregates
attention from the experts sends M-to-N traffic instead \cite{nr_zhu25}. On inference-shaped demand,
across expert-parallel degrees from 8 to 128 and packed without the port
condition, the schedule stays within 1\% of the classical lower
bound and never exceeds $11/9$, so the same cap on worst-case iteration
inflation, before switching delay, applies across that sweep. RL post-training inherits the
result through rollout, which is inference and takes from half to over
90\% of the step in reported systems \cite{ref41, nr_hu26b, nr_zhou25};
by Amdahl's law, a hypothetical 10\% rollout speed-up would be roughly a
5 to 9\% step speed-up. Weight synchronization between phases is a
separate load: under 1.6~s for a 72B model inside one cluster, against
rollouts of hundreds to thousands of seconds \cite{ref40}, but enough to
stall rollouts at terabyte scale, especially across data centers
\cite{nr_ye26b}.

\textbf{Availability and thermal are training-first:} Reported machine
failures run about $13.6\times$ lower per thousand machines per week for
an inference service than for pretraining at scale, 2.2 against 30
\cite{ref42}. If the per-laser rate fell by the same factor, the same
coverage bound would put a fixed four-laser reserve at 99.1\%
availability, against 86.3\% at
the stress training rate. The machinery pays off most where failures are
frequent and correlated, so we claim no per-component inference win for
LightReserve, ThermaLead, or TriageGuard. The case is strongest on the
largest jobs, since mean time to failure falls as the GPU count rises
\cite{ref35}; at platform scale, ByteDance logged 38,236 explicit and
5,948 implicit failures across 778,135 training jobs in three months
\cite{ref32}.

Disaggregated serving also moves the key-value cache from prefill to
decode over the fabric \cite{ref39, nr_patel24}, a load that grows with context. No component removes it, and we claim
no guarantee for it.

\subsection{Limitations and Open Problems}
\label{sect:limits}

The fallback rule of Section~\ref{sect:cpfrwk} covers the cases it can
detect: on a stale, low-confidence, or late signal, the component
bypasses itself. It does not catch a confident but wrong forecast, on
which a controller can reconfigure the fabric more often than it should.
Gating reconfiguration on confidence and feasibility curbs that churn
without eliminating it, and a formal bound on the harm is left to future
work. Acting also costs compute for the predictions and energy for the
heaters, and spoofed telemetry could force a mis-reconfiguration, so the
control interface has to be hardened.

Every numerical result outside Section~\ref{sect:hwimp} is a calibrated
simulation or a closed-form evaluation, and those of
Section~\ref{sect:hwimp} are synthesis, place-and-route, signoff, and
proof outputs, which settle the logic and its static timing but leave
the process and interfaces to fabrication; none is a measurement on
hardware. The two loss shares of the
goodput model are assumptions, and most of the stress-point gain
comes from provisioning the spare lasers that the coverage bound sizes
(Section~\ref{sect:corfaccomp}). Six limits bound the claims.

First, the failure inputs are assumed: the correlation $\rho$, on which
both the sparing result and the ceiling depend, and the per-laser
failure rate, for which we found no published field value for the
lasers inside external laser modules. The headline figures use the
stress setting $\lambda = 0.05$ per window, and 99.999\% stays out of
reach inside the $2\times$ cap of the 32-channel pool only while
$\lambda$ exceeds about $2.4 \times 10^{-4}$ (Table~\ref{tab:regime}).
We anchor $\rho$ in the common-cause beta factors of
Section~\ref{sect:analysis} but cannot yet calibrate it to field data
for these laser form factors, so the recommendation rests on the sweep
of Observation~\ref{obs:rhoinv}.

Second, of the three conditions that carry FoldFit's $11/9$ guarantee
onto a switch (Section~\ref{sect:ff}), two hold by construction here.
The third, realizing the packing as port-disjoint configurations, is
exact only when a demand may be split at a configuration boundary and
resumed later, and a proof for demands that must each run on one
uninterrupted circuit is open. The guarantee also covers dwell time
only, not the switching delay of each reconfiguration.

Third, ThermaLead's moving-average forecast does slightly worse than the dither
lock on the stress trace of Section~\ref{sect:tl}, and in the released
sweep its deficit grows to 49--67\% when the heat cycle's phase jumps
randomly every 16 steps or more often, and to 17\% under random
temperature steps of 1~\degC{} standard deviation every 4 steps. Gated, both
deficits stay within 4.3\%, an empirical limit, not a proof. A
deployable ThermaLead needs a forecast matched to the drift and aware of
regime changes.
Comparisons with the fastest reactive loops now reported
\cite{nr_shi26} and with heater pre-driving from kernel launches
\cite{nr_yoon26} remain open, since both need the control step in
seconds, which the per-step analysis leaves free.

Fourth, several external anchors are vendor-reported or forward-looking,
and each is hedged where it appears. Fifth,
Proposition~\ref{prop:quantile} covers only the correlation-driven part
of goodput, and Section~\ref{sect:tail} states what a genuine floor
under realized goodput would also require. Sixth, the composed goodput
turns on what a spare shortfall costs a real job, and the two readings
of Section~\ref{sect:corfaccomp} are cases, not bounds.

\subsection{The Measurement That Turns the Bound into a Forecast}
\label{sect:measure}

Theorem~\ref{thm:ceil} bounds availability in terms of $\rho$, so a
fleet-scale measurement of that correlation would pin down the value on
which the bound depends, and the measurement needs only instrumentation. Log the failure timestamp of every laser across many
identical pools, count how many lasers fail together in each restock window,
and estimate $\rho$ from how far that count is over-dispersed, that is,
how much more it varies than the independent-binomial expectation.
Random-effects lifetime models, which already estimate location-driven
correlation from GPU fleet failure data \cite{nr_clark26}, offer one way
to do it. Correlated failures are rare, so a credible estimate needs
exposure on the order of thousands of pool-years, from a hyperscaler
instrumenting its own packages or a vendor publishing co-failure
statistics. Two choices sharpen the result. Logging the candidate shared
causes lets $\rho$ be attributed to a mechanism. The full per-window
distribution should also be reported, because Part~(2) of
Theorem~\ref{thm:ceil} and Proposition~\ref{prop:quantile} show that
coverage and the goodput tail depend on the tail of $\Theta$, not only
on its variance. Until such data exists, $\rho = 0.15$ should be read as
a prior, and the design should rest on the full $\rho$-sweep. Cheaper
measurements would narrow other unmeasured inputs: one
instrumented package would ground the two composition loss shares, and
one real job would establish both the fabric's share of recoverable
stall and what a spare shortfall costs it.
\section{Conclusions}
\label{sect:conc}

This paper argues that each of the four optical layers can be run a
step ahead of failure. \textsf{Halley} puts one predictive controller over each
of them, though ThermaLead's simple forecast paid off mainly on smoother drift;
its gate kept it near the dither lock elsewhere. Each controller steps aside when its
forecast is flagged stale or unsure, so for such a forecast the worst
case is the reactive system it replaces; a confident but wrong forecast
has no such floor yet (Section~\ref{sect:limits}).

One theorem shapes the design: correlated laser failures put a ceiling on the coverage a large pool can
reach within a deployable budget cap. For a 32-channel pool at $\rho = 0.15$, 99.999\% stays out of reach
within the $2\times$ cap whenever a laser's failure probability per
restock window exceeds about $2.4 \times 10^{-4}$, as it does for a laser at the 50~FIT source target restocked once a
year. Between that target and the stress level, correlation raises the spares that 99.999\% takes more than tenfold. That target is then reached by decorrelating the
pool first and sizing the spares against the correlation that remains. Under the stated cost model, that path costs
\$44,100 at the stress operating point against \$234,000 for the pure-spares
alternative, a $5.3\times$ saving, and it stays the cheaper of the two
across the whole plausible band of $\rho$, which the field has not yet
measured.

Composed from the simulated components, effective cluster goodput at the
stress operating point rose 11.3 points on the coverage reading and 2.2
points on the lit-share reading. Most of the first figure comes from the
larger spare pool, and holding it to the $2\times$ cap leaves 11.1
points. The margin over today's reactive stack is narrowest under mild
conditions, about 1.3\% in relative terms at 50~FIT. How much of that
gain a real job would keep depends on the fabric's share of recoverable
stall, which this work did not measure. The part of the gain that a job
keeps reduces the accelerators and megawatts its training needs, by up to
3.3--14.7\% at the stress operating point. That much data center
capacity need not be built. The fallback that early action
relies on is synthesized, placed and routed for a shipping FPGA and
passes signoff as a 130~nm layout, and its combined core is machine-proven to match an independent specification
core on every cycle.

An operator should size the pool from a measured failure rate and a
measured correlation, spend the next dollar on separating the lasers
before buying more of them, and plan capacity against a bad window such
as the fifth percentile.

Further research can help realize a deployed system. What remains open,
ThermaLead's forecast and the theory, is stated in
Section~\ref{sect:limits}. Running these controllers on real optics
needs the parts of the telemetry and actuation interface that the optics
vendors have not yet published, chiefly per-ring temperature and lock
state. Estimating $\rho$ across a fleet needs only enough pools
instrumented for long enough.  This measurement would turn the ceiling
proved here from a bound conditioned on an assumed correlation into a
forecast.

\bibliographystyle{IEEEtran}
\bibliography{refs-oct4a}

%% \begin{IEEEbiography}[{\includegraphics[width=1in,height=1.25in,clip,keepaspectratio]{photos/Sooraj.jpg}}]
%% Biographies kept for the final version; commented out so the submission fits 25 pages.
%% \begin{IEEEbiography}[]
%%   {Manav Mishra\ } is a co-founder and CEO of Prompeteer.ai. His career
%%   runs from designing Graphics, Networking offload chips at Intel to
%%   helping found Microsoft's anti-spam, anti-phishing, team, through
%%   Google, successful Silicon Valley Startups, to leading AI and data
%%   products at MSCI. He holds several U.S. patents and is widely
%%   published. He is currently working on Agentic success for businesses.
  %% He received his M.S.  degree in Computer Science from Washington State
  %% University Pullman.
%%   \end{IEEEbiography}

%% \begin{IEEEbiography}[{\includegraphics[width=1in,height=1.25in,clip,keepaspectratio]{photos/KrishnaS.png}}]
%% \begin{IEEEbiography}[]
%% {Krishna M.  Sivalingam\ } holds the Gopalakrishnan Institute Chair
%% Professorship in the Indian Institute of Technology (IIT) Madras,
%% Chennai, INDIA. He received his Ph.D. and M.S.  degrees in Computer
%% Science from State University of New York at Buffalo in 1994 and 1990
%% respectively. Email: skrishnam@cse.iitm.ac.in
%% \end{IEEEbiography}

\end{document}